\documentclass[aps,prb,twocolumn,superscriptaddress]{revtex4-2}

\usepackage{amsmath}
\usepackage{amsfonts,amssymb, graphicx}
\usepackage{wrapfig}
\usepackage{afterpage}
\usepackage{natbib}
\usepackage{color}
\usepackage[usenames,dvipsnames,svgnames,table]{xcolor}
\usepackage{amsmath}
\usepackage{braket}
\usepackage{stmaryrd}

\usepackage{tikz}
\usepackage{pgffor}
\usepackage{verbatim}
\usepackage{float}
\usepackage{mathrsfs}
\usepackage{soul}
\usepackage[colorlinks,breaklinks=true,linkcolor=blue, citecolor=blue,urlcolor=blue, linktocpage=true]{hyperref}

\newcommand{\bpm}{\begin{pmatrix}}
\newcommand{\epm}{\end{pmatrix}}
\newcommand{\bmm}{\begin{matrix}}
\newcommand{\emm}{\end{matrix}}

\newcommand{\bea}{\begin{eqnarray}}
\newcommand{\eea}{\end{eqnarray}}

\newcommand{\heff}{\hbar_{\rm eff}}
\newcommand{\tilambda}{\tilde{\lambda}}
\newcommand{\Ccl}{C_{\rm cl}}

\newcommand{\llangle}{\left<\hspace{-5.83pt}\left<}
\newcommand{\rrangle}{\right>\hspace{-5.83pt}\right>}
\newcommand{\sllangle}{\left<\hspace{-4.83pt}\left<}
\newcommand{\srrangle}{\right>\hspace{-4.83pt}\right>}
\newcommand{\pluseq}{\hspace{3.5pt}^{^+}\hspace{-10.25pt}=}
\newcommand{\pluseqtext}{\hspace{3.5pt}^{^+}\hspace{-12pt}=\,}

\begin{document}

\newcommand{\fr}{\color{red}{}}

\title{Universal level statistics of the out-of-time-ordered operator}

\author{Efim B. Rozenbaum}
\email[]{efimroz@umd.edu}
 \affiliation{Joint Quantum Institute, University of Maryland, College Park, MD 20742, USA.}
\affiliation{Condensed Matter Theory Center, Department of Physics, University of Maryland, College Park, MD 20742, USA} 
\author{Sriram Ganeshan}
 \affiliation{Simons Center for Geometry and Physics, Stony Brook, NY 11794}
\affiliation{Department of Physics, City College, City University of New York, New York, NY 10031, USA}
\author{Victor Galitski}
 \affiliation{Joint Quantum Institute, University of Maryland, College Park, MD 20742, USA.}
\affiliation{Condensed Matter Theory Center, Department of Physics, University of Maryland, College Park, MD 20742, USA} 


\begin{abstract}
The out-of-time-ordered correlator has been proposed as an indicator of chaos in quantum systems due to its simple interpretation in the semiclassical limit. In particular, its rate of possible exponential growth at $\hbar \to 0$ is closely related to the classical Lyapunov exponent. Here we explore how this approach to quantum chaos relates to the random-matrix theoretical description. To do so, we introduce and study the  level statistics  of the logarithm of the out-of-time-ordered operator, $\hat{\Lambda}(t) = \ln \left( - \left[\hat{x}(t),\hat{p}_x(0) \right]^2 \right)/(2t)$, that we dub the ``Lyapunovian'' or ``Lyapunov operator'' for brevity. The Lyapunovian's level statistics is calculated explicitly for the quantum stadium billiard. It is shown that in the bulk of the filtered spectrum, this statistics perfectly aligns with the Wigner-Dyson distribution. One of the advantages of looking at the spectral statistics of this operator is that it has a well-defined semiclassical limit where it reduces to the matrix of uncorrelated classical finite-time Lyapunov exponents in a partitioned phase space. We provide a heuristic picture interpolating these two limits using Moyal quantum mechanics. Our results show that the Lyapunov operator may serve as a useful tool to characterize quantum chaos and in particular quantum-to-classical correspondence in chaotic systems, by connecting the semiclassical Lyapunov growth at early times, when the quantum effects are weak, to universal level repulsion that hinges on strong quantum interference effects. 
\end{abstract}

\maketitle

\section{Introduction}\label{sec:intro}
There exist a number of approaches to define the concept of ``quantum chaos.'' The basic approach is to quantize a classically chaotic model and declare the corresponding quantum model as ``quantum chaotic.'' Another prevailing method identifies quantum chaos with level repulsion between energy levels, described by the universal Wigner-Dyson statistics. The connection between the two is established via the so-called Bohigas-Giannoni-Schmit (BGS) conjecture~\cite{Bohigas84} (first formulated in Ref.~\cite{Casati80}), which postulates that the spectra of  time-reversal-invariant classically chaotic systems show the same fluctuation properties as predicted for a Gaussian orthogonal ensemble (GOE) of random matrices. Semiclassical approaches in the form of a periodic orbit theory~\cite{gutzwiller1991chaos} by Berry~\cite{berry1985semiclassical} and non-linear sigma models by Andreev {\it et al.}~\cite{andreev1996quantum, andreev1996semiclassical, altland2015review} have been employed to prove BGS conjecture with partial success. There are alternative approaches to quantum chaos: those based on wave-function behavior, such as quantum ergodicity~\cite{Stechel84, *Zelditch05, *Zelditch10}, Berry's random-wave conjecture~\cite{Berry77}, and nodal statistics~\cite{Berry02}; criteria based on transport or scattering properties~\cite{Smilansky88, *Casati90}; and definitions connecting to exponential behavior reminiscent of the classical instability, that is observed in quantum fidelity~\cite{Peres84, *Peres95}, Loschmidt echo~\cite{Wisniacki02,*Cucchietti02}, and out-of-time-ordered correlator (OTOC)~\cite{Rozenbaum17}. More recently, the ``definition'' of quantum chaos based on OTOC became the focus of much research due to its applicability to many-body quantum systems (see e.g. Refs.~\cite{kitaev, Maldacena16, *swingle2016measuring, *yao2016interferometric, *huang2016out, *fan2016out, *chen2016quantum, *swingle2016slow, *Syzranov17}). The quasiclassical limit of OTOC reproduces the sensitivity of quasiclassical trajectories to initial conditions. Exponential growth of OTOC at early times is identified as a fingerprint of quantum chaos, connecting the quantum dynamics to the hallmark of classical chaos -- the Lyapunov divergence of classical trajectories, colloquially known as the ``butterfly effect.'' 

In many cases (e.g., disordered metals~\cite{aleiner1996divergence, *aleiner1997divergence_2, *aleiner2016microscopic} and certain chaotic billiards) these approaches do appear equivalent, but there is no universal equivalence. For example, not all quantum models with Wigner-Dyson level statistics are required to have an  ``obvious" classical counterpart  (e.g., the Sachdev-Ye-Kitaev model~\cite{Sachdev15, kitaev}) and not all classically chaotic dynamical systems acquire Wigner-Dyson level statistics upon quantization, such as systems that show localization. Moreover, quantum systems with merely mixing (non-chaotic) classical counterparts can obey Wigner-Dyson distribution even without classical exponential instabilities (see, e.g., Ref.~\cite{Lima13}). Such cases are considered outside of the BGS characterization. This ambiguity makes the notion of quantum chaos somewhat poorly defined. It is highly desirable, therefore, to obtain a more straightforward way of connecting the different intuitive ideas and approaches to ``quantum chaos,'' and we attempt to do so in this work by introducing an operator, which we dub the Lyapunovian [see Eq.~(\ref{L}) below]. As we show, it contains information about both the development of universal level statistics -- resulting from quantum interference -- and classical Lyapunov exponents in a (semi)classical phase space.

Our study is motivated by  recent work on OTOCs~\cite{kitaev, Maldacena16, *swingle2016measuring, *yao2016interferometric, *huang2016out, *fan2016out, *chen2016quantum, *swingle2016slow, *Syzranov17}, the concept originally introduced by Larkin and Ovchinnikov~\cite{Larkin69} in the context of disordered metals. It involves a quantum expectation value of the following positive-definite operator:
\begin{equation}
\label{L}
\hat{C}(t) \equiv \exp{[ 2\,t\,\hat{\Lambda}(t)]}  = - \left[\hat{x}(t),\, \hat{p}_x(0) \right]^2,
\end{equation}
where we chose a pair of operators $\hat{x}(t)$ and $\hat{p}_x(t)$ -- the Heisenberg operators of a particle's $x-$coordinate and the corresponding component of its momentum. Both in the case of a dirty metal and a billiard, one can argue in the semiclassical limit that since $\hat{p}_x(0)=-i\hbar\frac{\partial}{\partial x(0)}$, the OTOC -- the quantum expectation value of the operator $\hat{C}(t)$ in Eq.~(\ref{L}) -- probes the sensitivity of quasiclassical trajectories to initial conditions: $C(t)\nobreak=\nobreak\langle\hat{C}(t)\rangle\nobreak=\nobreak\hbar^2\left\langle\left(\frac{\partial x(t)}{\partial x(0)}\right)^2\right\rangle$. Thus the classical Lyapunov-like growth is anticipated at early times, $C(t)\propto\exp(2\tilambda t)$, where $\tilambda$ is related to the classical Lyapunov exponent (see Sec.~\ref{sec:earlytime} for details).

However, whether the OTOC actually grows exponentially or not depends on the choice of a quantum state over which the expectation value is calculated. It also depends on the existence of a long enough time window within the Ehrenfest time scale $t<t_E$ (see Sec.~\ref{sec:earlytime}), before the quantum interference washes out the classical growth, if any. In some sense, the search for exponential growth of OTOC becomes the search for a quasiclassical description. In some cases, such as billiards or diffusive metals, the quasiclassical limit is obvious. In some others, such as the Sachdev-Ye-Kitaev model, the classical variables are ``hidden'' in the large-$N$ limit~\cite{maldacena2016remarks, bagrets2016sachdev}. The  dependence of the OTOC on the choice of a quantum state is a non-universal feature, and instead, motivated by Ref.~\cite{Cotler2017chaos}, we focus on the random-matrix structure of the Lyapunovian -- the Hermitian operator $\hat{\Lambda}(t)$ in Eq.~(\ref{L}). The Lyapunovian possesses a  semiclassical interpretation that enables us to connect the spectral statistics with that of the matrix of classical finite-time Lyapunov exponents in different cells of the partitioned phase space \footnote{Note the qualitative difference between the finite-time spectrum of the single-particle Lyapunovian and the spectrum of infinite-time Lyapunov exponents in multidimensional classical models~\cite{GurAri16,*Hanada18}.}. 

The rest of the paper is organized as follows. In Sec.~\ref{sec:model}, we introduce the specific model we used in the calculations. In Sec.~\ref{sec:univstat}, we demonstrate the main results on the universal level statistics of the Lyapunov operators. Next, Sec.~\ref{sec:tdls} elaborates on the dynamics of the time-dependent level statistics and the ways it can be observed. Sec.~\ref{sec:phasespace} gives a heuristic picture that helps in developing the intuition behind our findings. Finally, in Sec.~\ref{sec:earlytime}, we show the early-time exponential growth of OTOC in our model, and we explain why it is not always readily visible.

\begin{figure}
	\includegraphics[width=\linewidth]{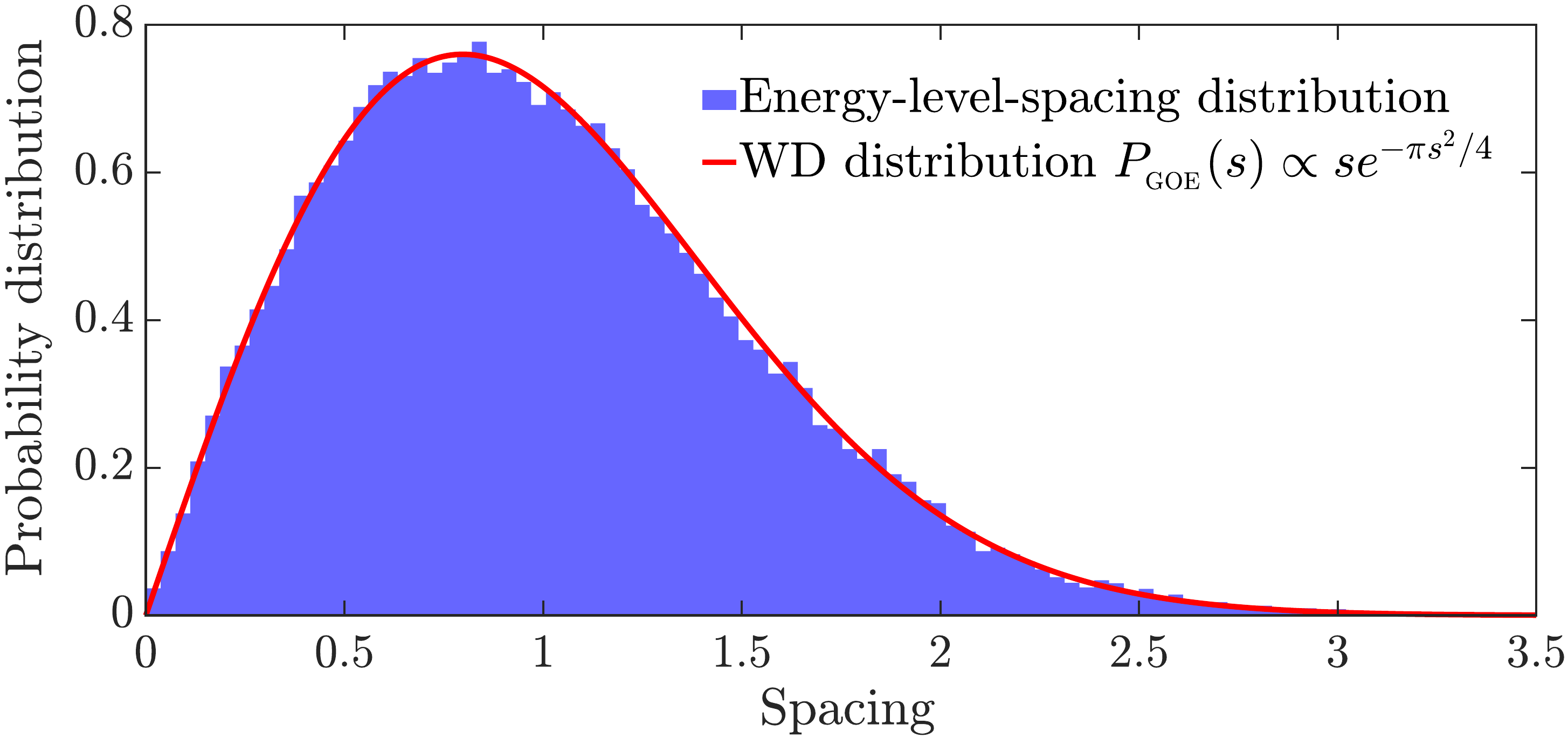}
	\caption{Energy-level statistics for quantum stadium billiard (separate for each eigenstate parity, combined~\cite{ParityComment}). Contribution from the bouncing-ball modes~\cite{Graf92, Sieber93, Alt99} is removed within the spectrum unfolding. Solid line shows GOE Wigner-Dyson distribution.} \label{fig:E_GOE} \vspace{-10pt}
\end{figure}

\section{Model}\label{sec:model}
For explicit calculations, we choose the quantum stadium billiard -- a canonical model to explore quantum signatures of chaos, -- but the main construction naturally transplants to a wide class of models. The classical Bunimovich stadium billiard~\cite{Sinai70, Bunimovich74, *Bunimovich79, *Bunimovich91, Benettin78, Dellago95, Biham92} is a seminal model of classical chaos, and its quantum counterpart has been known  to obey the Wigner-Dyson energy-level statistics of GOE~\cite{McDonald79, Casati80, Bohigas84, Shudo90, Borgonovi96_1, Alt99} reproduced in Fig.~\ref{fig:E_GOE}. The oscillatory contribution of the bouncing-ball orbits~\cite{Graf92, Sieber93, Alt99} to the density of states -- a non-generic feature of the stadium -- is subtracted in order to obtain the near-perfect agreement between the level-spacing distribution and the Wigner surmise. Throughout the paper, we consider the billiard with unit aspect ratio $a/R=1$,  where $2a$ is the length of the straight segments of the walls and $R$ is the radius of the circular ones. We use the units where both the area of the billiard $A=(\pi+4)R^2$ and the particle mass $m$ are set to $1$. We also choose a certain momentum $p_0$ as the third unit. Later, it will play the role of the quantum-particle's average momentum. In the semiclassical limit, $p_0$ translates into the momentum of the classical particle inside the billiard. In these units, the Schr\"{o}dinger equation and the boundary condition read:
\begin{equation}
\label{SE}
\hspace{-0.7pt} -\dfrac{\heff^2}{2}\nabla^2 \Psi(x,y) = E\Psi(x,y),\,\,\  \Psi({\bf r})\Bigl|_{{\bf r}\in {\rm billard\, walls}} \equiv 0,
\end{equation}
where $\heff=\hbar/(p_0\sqrt{A})$. The stadium billiard has two reflection symmetries: $x\leftrightarrow -x$ and $y\leftrightarrow -y$. Correspondingly, its eigenstates have one of four possible parities~\cite{McDonald79}. E.g., the odd-odd-parity functions $\Psi_{\rm oo}(-x,y)\equiv\Psi_{\rm oo}(x,-y)\equiv -\Psi_{\rm oo}(x,y)$. As is usually done, in order to enforce these parities and speed up the calculations, we use a quarter of the billiard imposing Dirichlet and/or Neumann boundary conditions on the cuts to obtain solutions of all four parities separately. 

We solve these boundary-value problems for the Laplace operator numerically using the finite-element method. It is known that the accuracy of the numerical solution deteriorates with the number of found eigenstates~\cite{Heuveline03}. We use Weyl's formula for the number of modes~\cite{Baltes78} to control it. According to Weyl's law, the average number of eigenstates below energy $E$ asymptotes to: 
\begin{equation}\label{eq:weyl}
\mathcal{N}(E)\simeq \frac{A}{4\pi}\frac{2}{\heff^2}E-\frac{P}{4\pi}\sqrt{\frac{2}{\heff^2}E},\quad E\to\infty, 
\end{equation}
where $P$ is the billiard's perimeter. We do all calculations in several ranges. The smallest range is limited to about $N=5000$ eigenstates and preserves almost exact agreement with Weyl's formula, and the largest one is over $N=10^5$ states. We verify that our results do not depend on the truncation size $N$. In addition, we benchmark our solutions against those we obtain independently via the boundary-integral method, and we reach the same level of accuracy with both approaches. We should note that the absolute error in the number of found energy levels (as compared to Weyl's formula) grows quadratically with energy for the levels $E_n,\; n \gtrsim 2000$ with a very small prefactor. However, while the overall magnitude of the energy starts to overestimate Weyl's expression -- the inverse of Eq.~(\ref{eq:weyl}) -- after this point, the structure of the spectrum is preserved. This is verified by varying the algorithm's accuracy, comparing the results to those obtained via the boundary-integral method, and subtracting the smooth quadratic function that brings the spectra obtained by all methods on top of each other. In the tests we performed, our results for the distributions did not show any influence of this deviation as it is completely canceled by the spectrum unfolding anyway.

\section{Universal statistics of the Lyapunovian}\label{sec:univstat}
Let us turn to the central subject of the work -- the level statistics of the out-of-time-ordered operators. Apart from the Lyapunovian [Eq.~(\ref{L})], we also define the Hermitian operators: \vspace{-3pt}
\begin{equation}
\label{Lk}
\hat{C}^{(k)}(t) = (-i)^k \left[\hat{x}(t), \hat{p}_x(0) \right]^k \pluseq \exp{[k\,t\, \hat{\Lambda}_k(t)]},
\end{equation} 
with $k\in\mathbb{N}$, such that $\hat{C}^{(2)}(t)\equiv\hat{C}(t)$. For even $k=2n$, $\hat{\Lambda}_{2n}(t)\equiv\hat{\Lambda}(t)$, while for odd $k=2n-1$, we only define $\hat{\Lambda}_{2n-1}(t)$ within the positive-eigenvalue subspaces of $\hat{C}_{2n-1}(t)$, which is indicated by the  ``$\pluseqtext$'' sign. In addition, we consider a closely related Hermitian operator that defines a four-point-correlator part of OTOC:
\begin{equation}
\label{Lt}
\hat{F}(t) = \hat{x}(t)\hat{p}_x(0)\hat{x}(t)\hat{p}_x(0) + {\rm H.~c.} \pluseq \exp{[ \hat{\Gamma}(t)]}.
\end{equation}

We use the energy eigenstates $\ket{E_n}$ to construct matrices ${C}^{(k)}_{nm}(t)=\braket{E_n|\hat{C}^{(k)}(t)|E_m}$ and $F_{nm}(t)=\braket{E_n|\hat{F}(t)|E_m}$. For numerical calculations, we truncate the operators to finite $N\times N$ matrices according to the number of eigenstates in use. Then the finite matrices are numerically diagonalized and the statistics of the spacings between the logarithms of eigenvalues as well as between the eigenvalues themselves are studied. Due to the definite parities of the energy eigenfunctions, the matrices $C_{nm}^{(k)}(t)$ and $F_{nm}(t)$ are $4\times4$ block-diagonal, and each block corresponds to one parity. Level spacings are thus only calculated within each block separately (because eigenvalues in different blocks are not correlated with each other), and then these four sets of spacings are combined for statistical analysis. The operators $\hat{C}^{(k)}$ and $\hat{F}$ have the same bulk level statistics as their respective logarithms, $\hat{\Lambda}_k$ and $\hat{\Gamma}$~\cite{WishartNote}. Therefore, we only show the results for the logarithmic operators. We observe different ensembles for different operators. 

\begin{figure} 
	\includegraphics[width=\linewidth]{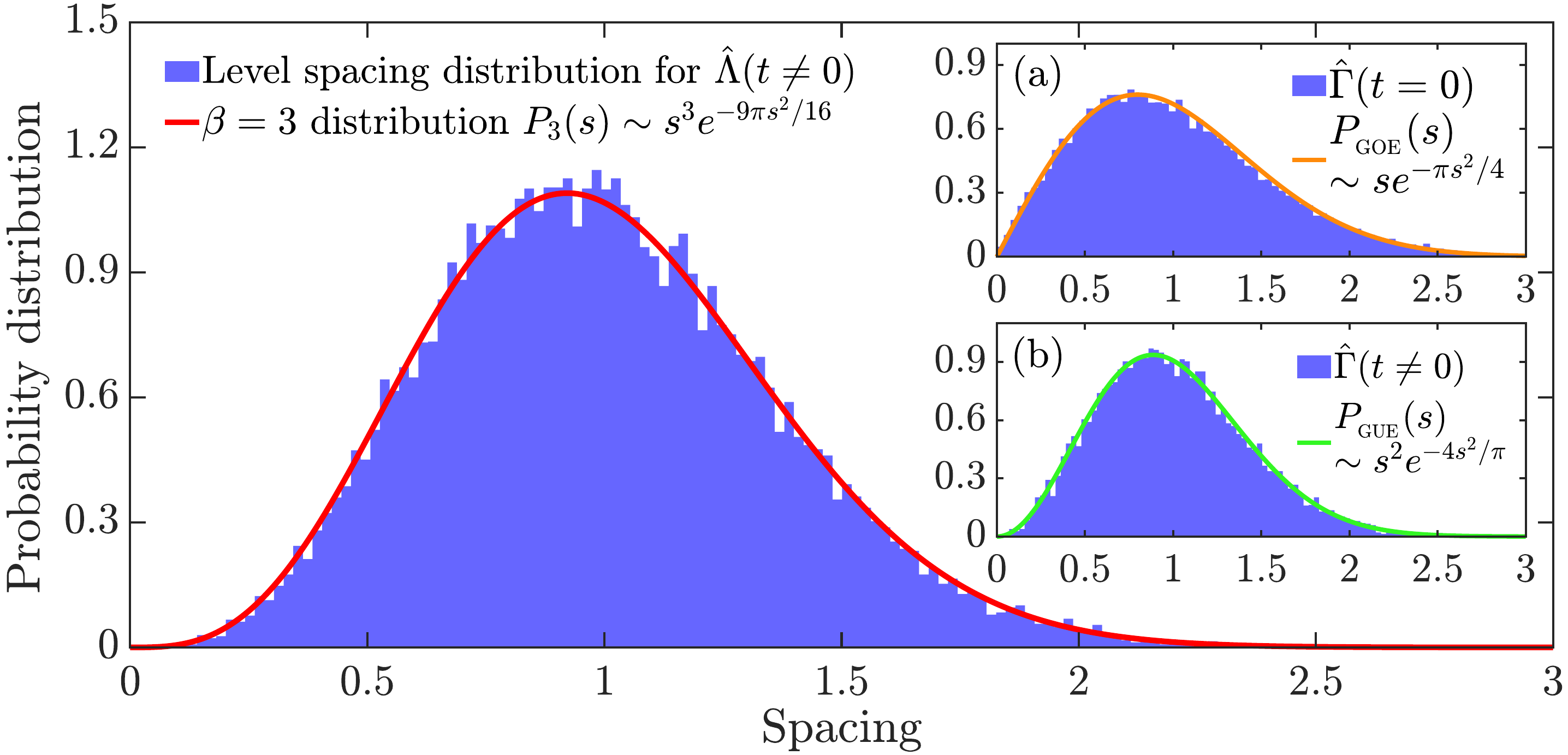}
	\caption{{Eigenvalue-spacing distribution for the bulk of the Lyapunovian spectrum for every second state (within each parity block, combined). The total number of levels is $10^5$. Insets: (a) bulk level spacing distribution for $\hat{\Gamma}(t=0)$; (b) the same for $\hat{\Gamma}(t\neq0)$. Solid lines show the corresponding Wigner-Dyson distributions.}} \vspace{-10pt} \label{fig:GUE_GSE}
\end{figure}
Note that at $t=0,\;$  $\hat{C}^{(k)}(0)=\heff^k$ are $c$-numbers, so they do not have level-spacing distributions. However, the operator $\hat{F}(0)=\hat{x}(0)\hat{p}_x(0)\hat{x}(0)\hat{p}_x(0)+{\rm H.~c.}$ is a non-trivial Hermitian operator, and its matrix $F_{nm}(0)$ is real-valued. We find -- see inset (a) in Fig.~\ref{fig:GUE_GSE} -- that the bulk level statistics for $\hat{\Gamma}(0)$ [and for $\hat{F}(0)$] corresponds to GOE -- the same ensemble as that of the Hamiltonian. The reason for this can be understood by representing the momentum operator as $\hat{p}_x = \frac{i}{\heff}[\hat{H},\hat{x}]$, where $\hat{H} = \frac{\hat{p}_x^2+\hat{p}_y^2}{2}+V_{\rm walls}(\hat{x},\hat{y})$ is the Hamiltonian of the billiard. Then $\hat{F}(0) = -\heff^{-2}\left(\hat{x}\,[\hat{H},\hat{x}]\right)^2 + {\rm H. c.}$

At any finite time, $t\neq0$, all $C^{(k)}_{nm}(t)$ and $F_{nm}(t)$ become non-trivial Hermitian matrices with complex entries due to the unitary evolution of the operator $\hat{x}(t)=e^{i\hat{H}t}\hat{x}e^{-i\hat{H}t}$ with the random-matrix-like Hamiltonian. In Fig.~\ref{fig:GUE_GSE}, main plot and inset (b) show the bulk level statistics of the Lyapunovian and $\hat{\Gamma}(t)$, respectively, at a fixed time $t\neq0$. Of course, microscopic details of both spectra are different and time-dependent, as the individual eigenvalues move with time. But we find that their bulk spectral statistics appear to be completely universal and remain the same for any $t\neq0$. We should stress that for the operators defined this way in the entire Hilbert space, there is no notion of short time (such as the collision or Ehrenfest times), so all times are equivalent, indeed. On the other hand, as shown in Sec.~\ref{sec:tdls}, one can observe dynamical evolution of the spectral properties of these operators when they are projected to a sub-space of the Hilbert space that consists of initially non-overlapping classical-like states only. In this case, after these classical-like states ``dissolve'' in the semiclassical phase space as the time reaches and exceeds $t_E$, the statistics tends to develop from the initial uncorrelated Poisson-like one to the Wigner-Dyson statistics -- similar to that shown for the operators in the entire Hilbert space in Fig.~\ref{fig:GUE_GSE}.

The bulk level statistics of $\hat{\Lambda}_{2n-1}$ and $\hat{\Gamma}$ correspond to GUE [Fig.~\ref{fig:GUE_GSE}, inset (b)], while extracting level statistics of the Lyapunovian [the operators $\hat{\Lambda}\equiv\hat{\Lambda}_{2n}$ and $\hat{C}^{(2n)}$] requires one more step. The bulk level statistics of $\hat{C}^{(1)}$ and $\hat{\Lambda}_1$ correspond to GUE. But since the spectrum of $\hat{C}^{(1)}$ has positive and negative branches, and $\hat{C}\equiv\left[\hat{C}^{(1)}\right]^2$, the spectrum of $\hat{C}$ consists of these positive and negative branches squared and superimposed onto each other (this translates to the spectrum of the operator $\hat{\Lambda}$, as well). This results in the effective suppression of level repulsion, because the neighboring levels that originate from different branches of the spectrum of $\hat{C}^{(1)}$ have no short-range correlation. We present two ways to account for this effect. First, provided the knowledge of the spectrum of $\hat{C}^{(1)}$, one can filter the eigenvalues of $\hat{C}$ that originate from only one -- positive or negative -- branch. This results in the GUE filtered bulk level statistics for $\hat{C}$ and the Lyapunovian. Alternatively, without the knowledge of the spectrum of $\hat{C}^{(1)}$, but given that it is approximately evenly distributed around zero (the matrix tends to be traceless as its size is increased), one can filter every second eigenvalue of $\hat{C}$ to greatly reduce the fraction of uncorrelated neighboring eigenstates. Following this approach, for every second level in the bulk of the spectra of $\hat{C}$ and $\hat{\Lambda}$, one finds the Wigner-Dyson distribution that corresponds to the Gaussian ensemble with the Dyson index $\beta=3$ -- intermediate between GUE and GSE [Fig.~\ref{fig:GUE_GSE}, main plot].

While the former (GUE) result is natural, the $\beta=3$ ensemble for every second level of the Lyapunovian results from the combination of the operator's intrinsic structure and the filtering algorithm. However, it is still general -- the same statistical properties can be found for next-nearest-neighbor level spacing in the bulk of the spectra of positive-definite matrices of the form $M^2$ (or $\ln M^2$), where $M$ is an Hermitian random matrix drawn from GUE. This argument suggests that for all odd powers $2n-1$, the bulk level statistics of $\hat{C}^{(2n-1)}(t\neq0)$ should correspond to GUE, and for all even powers $2n$, the bulk level statistics for every second level of $\hat{C}^{(2n)}(t\neq0)$ should correspond to the Gaussian ensemble with $\beta=3$. We have verified that it is indeed the case for $k=1,2,3,\mbox{and }4$. 

We stress that in integrable models, the spectral structure of the Lyapunovian-type operators is drastically different from that in the non-integrable ones. There are multiple degeneracies in the Lyapunov-operator spectra in the integrable case, and the corresponding level-spacing distributions of the operators $\hat{\Lambda}_1$ and $\hat{\Lambda}$ are thus very tightly peaked around zero and are generally not even well defined, because the unfolding procedure cannot be performed. We checked it specifically for a circular billiard, a rectangular billiard, and for a 1D particle-in-a-box model (semi-analytically). In all these cases, the level repulsion is absent, and most of the Lyapunovian eigenstates are (quasi)degenerate. So, one can readily distinguish such systems from the chaotic ones.

\begin{figure} 
	\includegraphics[width=0.9\linewidth]{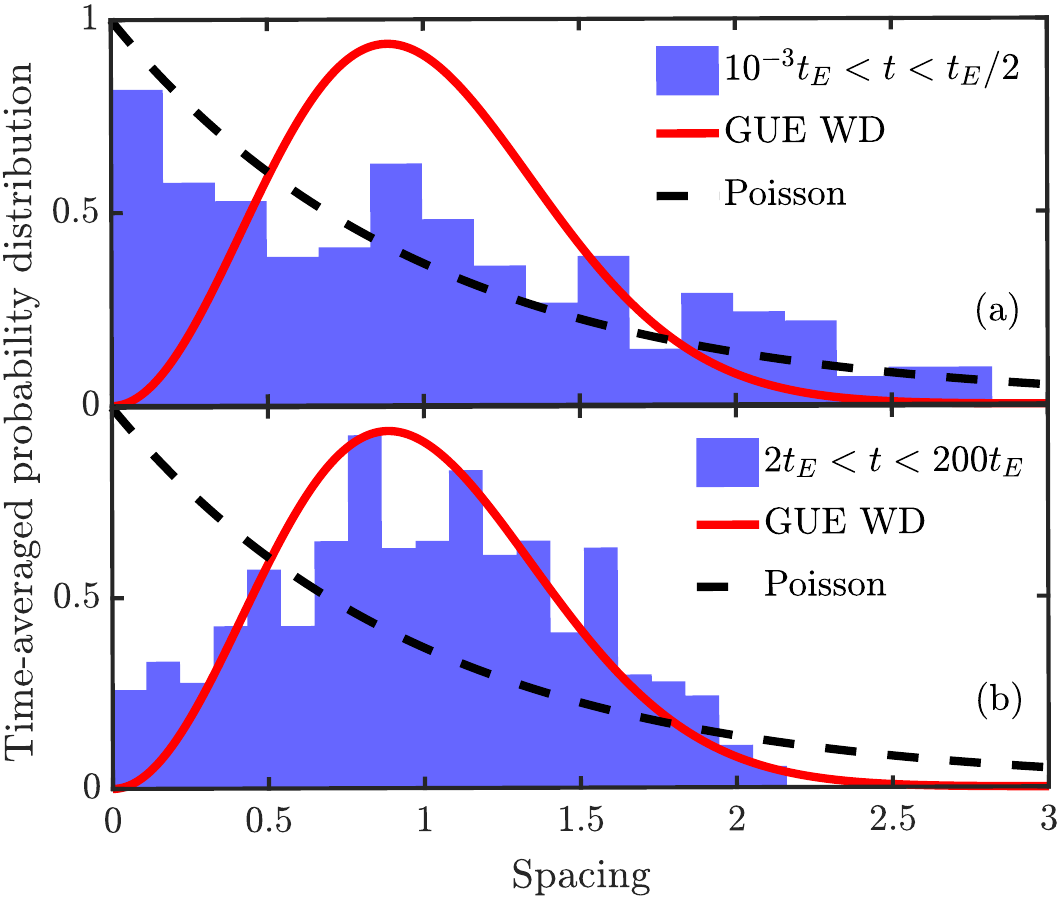}
	\caption{Eigenvalue-spacing distribution for the bulk of the spectra of an ensemble of projections of $\hat{C}^{(1)}(t)$ onto the coherent-state subspaces averaged over that ensemble and over time in two ranges of time: (a) at $t<t_E$, the distribution shows clear signatures of the Poisson component related to the uncorrelated nature of the phase space; (b) at $t>t_E$, the statistics tends to the universal GUE Wigner-Dyson distribution as phase-space correlations build up. With larger matrices, one can see that it becomes exact, such as the one shown in Fig.~2(b) in the main text. The low quality of the histograms is related to the small size of the subspaces ($8\times8$ matrices).} \label{fig:Poisson_to_WD}
\end{figure}

\section{Time-dependent level statistics}\label{sec:tdls}
We now turn to the particularly interesting question of the connection between the exponential Lyapunov growth of the OTOC, $C(t)=\braket{\Psi|e^{2\,t\,\hat{\Lambda}(t)}|\Psi}\propto e^{2\tilambda t}$, at early times and the Wigner-Dyson level statistics of the operator $\hat{\Lambda}(t)$. There appears to be a disconnect between the two: the former -- the Lyapunov growth -- is an early-time ($t<t_E$) classical behavior in the absence of quantum interference, while the latter is a consequence of well-developed quantum interference. We begin with a schematic demonstration of the mechanism of the correlation buildup between initially almost uncorrelated classical-like states. It also translates to the correlation build-up between the phase-space cells discussed in the next section. We start by projecting the operator $\hat{C}^{(1)}(t)$ onto an ensemble of 20 subspaces of the Hilbert space to form an ensemble of 20 projected operators (to improve statistics). Every subspace is composed of 8 almost non-overlapping minimal-uncertainty wave packets; each has unit average momentum. Note that although \emph{all} possible coherent states form an over-complete basis, we do not have to project operators onto all of them and, instead, have to take a subset that consists of states that form an (almost) orthonormal basis in the corresponding subspace. Our subsets that satisfy these requirements are small due to numerical limitations, but in principle they can be arbitrarily large, given small enough $\heff$. Letting these states evolve in time, we calculate the eigenvalue-spacing distribution for the projected operators at different times (excluding the smallest and the largest eigenvalues). Then we average these distributions over the ensemble of projected operators and, for better statistics, over time in two intervals: short times (between $10^{-3}t_E$ and $t_E/2$) and long times (between $2t_E$ and $200t_E$). After unfolding, we obtain distributions that roughly show the conversion from the uncorrelated -- Fig.~\ref{fig:Poisson_to_WD} (a) -- to the correlated -- Fig.~\ref{fig:Poisson_to_WD} (b) -- state of the phase space. The quality of the distribution is very limited by the small number of non-overlapping classical-like states that we fit into the billiard, but the principle can be observed.

\section{Phase-space description of OTOC}\label{sec:phasespace}
We study two related phenomena: (i) the exponential growth of OTOC at early times (to be discussed in Sec.~\ref{sec:earlytime}) and (ii) the transition in the level statistics of ``the projected Lyapunovian'' from the Poisson to the Wigner-Dyson distribution. Here ``the projected Lyapunovian'' is a shorthand referral to a projection of the Lyapunov operator to a subspace of virtually non-overlapping classical-like states, as discussed above. To develop further intuition about the connection between (i) and (ii), we follow Cotler {\it et al.}~\cite{cotler2017out} and consider the Lyapunov operator within the phase-space formulation. This is achieved by describing the quantum dynamics in terms of the Wigner function, $W({\bf r},{\bf p},t)$, in the four-dimensional phase space that we parametrize by $z=({\bf r},{\bf p})$ for brevity. All operators are translated into phase-space distributions via the Wigner transform~\cite{Moyal49}.

In particular, the out-of-time-ordered operator $\hat{C}(t)$ corresponds to the Moyal brackets: 
\begin{equation}
C_{\rm MB}(z,t)=-\llbracket X(z,t),P(z,0)\rrbracket^2,
\end{equation}
where we can choose $P(z,0)=p_{\rm cl}(z,0)$ to be classical, and $X(z,t)$ is the solution of the Moyal evolution equation: $\dot{X}(z,t)=\llbracket H(z),X(z,t)\rrbracket$, where we also choose a classical initial condition $X(z,0)=x_{\rm cl}(z,0)$. These choices correspond to the projection we introduced in the previous section. We can then express 
\begin{equation}
X(z,t)=x_{\rm cl}(z,t)+\sum^{\infty}_{k=1}\heff^{2k}x^{(2k)}(z,t),
\end{equation}
and the series of quantum corrections vanishes at $t=0$ according to the initial conditions: $x^{(2k)}(z,0)=0$. This choice of initial conditions ensures that $X(z,t)$ is the Moyal trajectory which coincides with the classical trajectory $x_{\rm cl}(z,t)$ in the $\heff\to0$ limit. The classical trajectories are obtained by solving the Hamilton-Jacobi equation. The $\heff$-dependent corrections are obtained by solving the series of the following evolution equations:
\begin{align}
\dot x^{(2n)}(z,t)&=\sum^n_{k=0}\llbracket H(z), x^{(2k)}(z,t) \rrbracket_{2(n-k)},
\end{align}
where the indexed brackets are defined as 
\begin{equation}
\llbracket A,B\rrbracket_{2n}\equiv\frac{A(z)\left( \overleftarrow{\partial}_{\bf r}\overrightarrow{\partial}_{\bf p}-\overleftarrow{\partial}_{\bf p}\overrightarrow{\partial}_{\bf r}\right)^{2n+1}B(z)}{\left(2n+1\right)!\, \left(-4\right)^n}. 
\end{equation}
The initial conditions for the higher-order corrections are $x^{(2k)}(z,0)=0$ for all $k>0$, since at time $t=0$ all distributions are classical and are captured within the Poisson-bracket term of the evolution equation. 

\begin{figure} 
	\includegraphics[width=\linewidth]{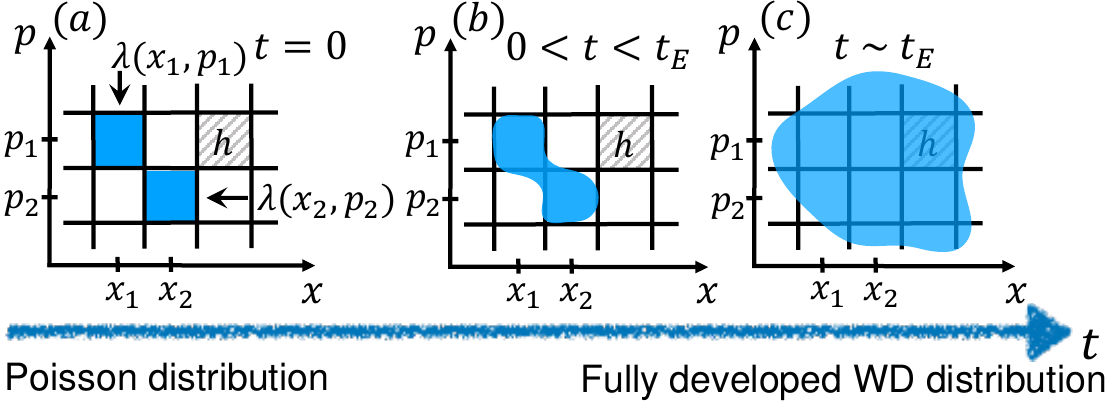}
	\caption{Schematics of the correlation development in phase space with time if initial states are semiclassical. (a) At times $t \ll t_E$, the local finite-time Lyapunov exponents are independent in different cells. (b) As time goes towards $t_E$, the correlations build up. (c) Around $t_E$, the phase-space becomes fully correlated, as shown by the distributions in Fig.~\ref{fig:GUE_GSE}.} \vspace{-10pt} \label{fig:StatvsTime}
\end{figure}

In this semiclassical approach, the classical phase space can be thought of as partitioned into the cells with the phase volume $\delta z=(2\pi\heff)^2$. Within the phase-space formulation, the Lyapunov operator is represented via a matrix whose indices enumerate these cells. The elements of this matrix are functions supported only within one cell. The $\heff$-expansion of the corresponding evolution shows that the zeroth-order Larkin-Ovchinnikov classical term, $[\partial x_{\rm cl}(z,t)/\partial x(z,0)]^2\propto e^{2\lambda(z)t}$, leads to independent Lyapunov exponents for each cell [Fig.~\ref{fig:StatvsTime}(a)]. In other words, the Lyapunov operator in the classical limit is a matrix of uncorrelated Lyapunov exponents.  A typical correlation term comes from an expression of the type $\heff^2\left[\partial x^{(2)}(z,t)/\partial x(z,0)\right]\left[\partial x_{\rm cl}(z,t)/\partial x(z,0)\right]$, which is the $\heff^2$-order correction to the trajectory~\cite{cotler2017out}. The $\heff^2$-dependent corrections to $C_{\rm MB}(z,t)$ generate correlations between the cells, and repulsion between the eigenvalues of the Lyapunov matrix ``commences'' [Fig.~\ref{fig:StatvsTime}(b)]. Such correlations fully develop around the Ehrenfest time when the phase space becomes highly correlated~\cite{zurek1994decoherence} leading to the breakdown of the Moyal expansion -- or any semiclassical description of OTOC~\cite{Larkin69} [Fig.~\ref{fig:StatvsTime}(c)]. The full quantum operators such as $\hat{C}(t)$ generally correspond to late times ($t>t_E$) in this picture, since they encapsulate full quantum interference effects resulting in the universal Wigner-Dyson statistics as shown in Fig.~\ref{fig:GUE_GSE}. However, as shown in Sec.~\ref{sec:tdls}, when these operators are projected to a subspace of initially classical-like states, their eigenvalue-spacing statistics changes across the Ehrenfest time from the Poisson-dominated distribution to the Wigner-Dyson one.

\section{Early-time behavior of OTOC}\label{sec:earlytime}
Finally, we address the question of how to actually extract the classical Lyapunov exponent from the Lyapunov operator in a way similar to that in Ref.~\cite{Rozenbaum17}. As noted above, not every matrix element would result in the exponential growth. For example, Hashimoto {\it et al.}~\cite{Hashimoto17} reported a lack of exponential growth in the thermal average of the out-of-time-ordered operator -- defined as $\mbox{OTOC}_\beta(t)=Z^{-1}\sum\limits_n e^{-\beta E_n}\braket{E_n|\hat{C}(t)|E_n}$ -- for the quantum stadium billiard. One would expect it to be the case, indeed, because the quantum thermal state in this system has no semiclassical description, which would correspond to a particle moving with a definite velocity. Instead, it mixes up different momenta and positions. So, this thermal average involves the states with well-developed quantum interference, where no classical dynamics is present already at $t=0$. In addition, it primarily accounts for ``the most quantum'' low-energy states (unless the temperature $\beta^{-1}$ is very high) that also have low momenta, while the Lyapunov exponent is proportional to the momentum. 
\begin{figure} 
	\includegraphics[width=\linewidth]{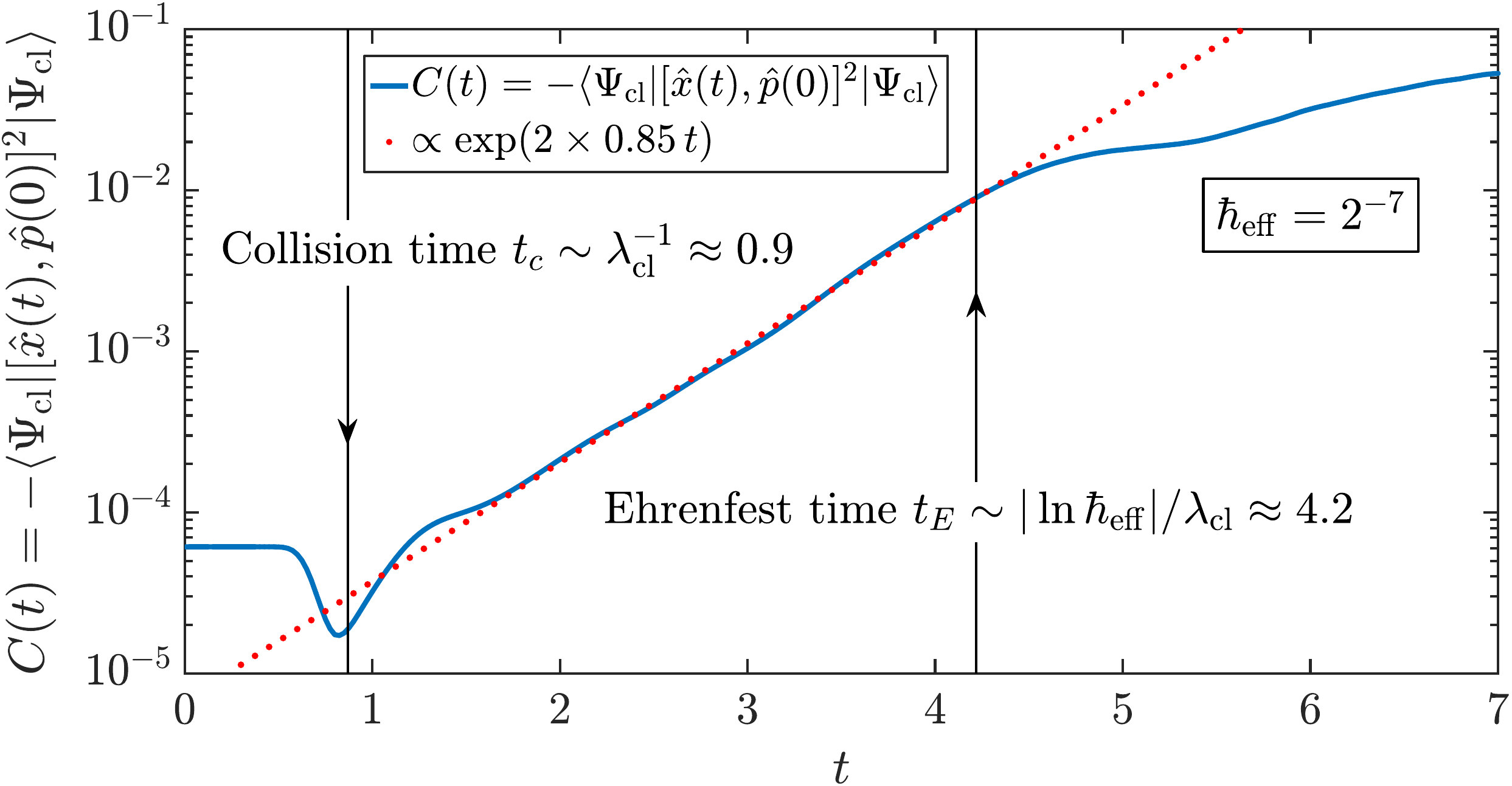}
	\caption{OTOC as the operator~(\ref{L}) averaged over the initial state~(\ref{eq:InitialState}) at early times (semi-log scale). $\heff=2^{-7},\;x_0=y_0=0,\;p_{0x}/p_{0y}=e,\;\sigma=1/\sqrt{2}$. Between $t_c$ and $t_E$, the growth is nearly exponential, $C(t)\propto e^{2\tilambda t}$, for time longer than $4/(2\tilambda)$, but the value of $\tilambda$ is not self-averaged yet.} \label{fig:OTOC_earlytime} \vspace{-10pt}
\end{figure}

To achieve exponential growth in this and, we believe, in many other systems, we have to identify ``the most classical'' initial state and let it evolve with time. In the case of a billiard, the natural choice  is a Gaussian minimal-uncertainty wave packet:
\begin{equation} \label{eq:InitialState}
\Psi_{\rm cl}({\bf r}) \propto \exp{ \left[ -\frac{ ({\bf r} - {\bf r}_0)^2}{2\heff\sigma^2} + \frac{i}{\heff} {\bf p_0}\cdot {\bf r} \right]},
\end{equation}
where $\sigma$ controls initial squeezing, and the parameters ${\bf r}_0$ and  ${\bf p_0}$ are the initial average position and momentum of the wave packet. $|{\bf p}_0|=p_0=1$ is the unit of momentum introduced before. 

Let $\lambda_{\rm cl}$ denote the classical Lyapunov exponent of the system at unit momentum $|{\bf p}|=1$ (the mass is fixed at $m=1$, so $\lambda_{\rm cl}|_{\bf p}\propto |{\bf p}|$). There are two relevant time scales: the collision time $t_c\sim1/\lambda_{\rm cl}$ is of the order of the time it takes the wave packet to hit the billiard's wall, and the Ehrenfest time $t_E\sim|\ln\heff|/\lambda_{\rm cl}$ is of the order of the time it takes a minimal-uncertainty wave packet to spread across the entire system. Classically, $\lambda_{\rm cl}$ is defined as the infinite-time average and it can be obtained for (almost) any initial condition by allowing enough time for a trial trajectory to explore a sufficient fraction of the phase space. At early times, though, the exponent fluctuates a lot before it reaches its average value, and the early-time values depend on the initial conditions. In the quantum calculation, the classical physics is limited to $t<t_E$, which in our case allows for just a few collisions with the walls. But instead of a single trial trajectory, we start with a wave packet that is equivalent to averaging over an ensemble of trajectories, which, in turn, is equivalent to averaging over a longer time and decreases the fluctuations. Within our numerics, we were still unable to reach complete self-averaging, so while we see a robust exponential growth spanning the interval between $t_c$ and $t_E$, the value of the exponent does depend on the initial wave packet and fluctuates moderately. However, it does not indicate any disagreement between quantum and classical description at early times. Classically, one can see the same fluctuations in the short-time Lyapunov exponent averaged over Wigner distributions of initial conditions that correspond to minimal-uncertainty wave packets used as initial conditions in our quantum calculations \cite{Crt}. The fluctuations occur both as functions of time and initial conditions.

As shown in Fig.~\ref{fig:OTOC_earlytime}, at early times ($t<t_E$), OTOC does grow exponentially: $C(t)\propto e^{2\tilambda t}$. In this semiclassical regime, we can replace the commutator with the Poisson brackets and average them classically over the ensemble of trajectories that corresponds to the Gaussian Wigner distribution $W_{\rm cl}(z)$ built from the initial state $\ket{\Psi_{\rm cl}}$. We denote this average as $\,\llangle\;\dots\;\rrangle\,$~\cite{ProbDist}. We then have $C(t)\approx\Ccl(t)$ at $t<t_E$, where:
\begin{align}\label{eq:QuantumOTOC}
C(t) &= \braket{\Psi_{\rm cl}|\hat{C}(t) |\Psi_{\rm cl}} \propto e^{2\tilambda t}, \\ \label{eq:ClassOTOC}
\hspace{-2pt} \Ccl(t) &= \hbar^2\llangle \hspace{-3pt} \left(\frac{\partial x(z,t)}{\partial x(z,0)}\right)^2 \rrangle \propto \sllangle \hspace{-1pt} e^{2\lambda_{\rm cl}^{\rm ps}(z, t) t}\srrangle = e^{2\lambda t},\hspace{-3pt}
\end{align}
and $\lambda_{\rm cl}^{\rm ps}(z,t)$ accounts for both the proportionality to the total momentum and the short-time effects giving $\lambda_{\rm cl}^{\rm ps}(z,t)$ the dependence on the rest of the phase-space coordinates and time. Note that $\lambda$ in Eq.~(\ref{eq:ClassOTOC}) is very close in spirit to the notion of the expansion entropy used for the recently updated definition of classical chaos~\cite{Ott15}. Strictly speaking, one has to compare the quantum exponent $\tilambda$ to the classical value of $\lambda$. But as noted above, available time $t<t_E$ is not sufficient for the quantum exponent $\tilambda$ to self-average, and we do not reach exact quantitative agreement. Instead, in various calculations, we got $\tilambda$ in the interval between $\lambda_{\rm cl}/2$ and $3\lambda_{\rm cl}/2$, while $\lambda\gtrapprox\lambda_{\rm cl}$. $\lambda_{\rm cl}\approx1.15$ is calculated for the classical stadium billiard in Refs.~\cite{Benettin78,*Dellago95,*Biham92,LE_Comment}. We reproduced the same value in our classical-billiard calculation. The example of the quantum-billiard calculation in Fig.~4 has $\tilambda\approx0.85$ \footnote{Note that classical Lyapunov exponent was first extracted in a related way in Refs.~\cite{Wisniacki02,*Cucchietti02} from Loschmidt echo, which is closely related to OTOC.}.

\section{Summary and Outlook}\label{sec:summary}
We proposed a tool to study and even define quantum chaos in general quantum systems -- the Lyapunov operator. We investigated the behavior of its level statistics and benchmarked it in a simple example of the stadium billiard. The Lyapunovian allowed us to unify the early-time signatures of chaos in the absence of quantum interference and the late-time ones related to well-developed interference in a single quantity. Moreover, the Lyapunov operator can probe the transition between the two regimes and generalize a straightforward intuition behind the quantum-to-classical correspondence to a wider class of quantum systems. As compared to the OTOC, the Lyapunovian is free from the ambiguity of the initial-state choice, and thus it can more reliably answer the question of regular-vs-chaotic nature of a given system.

We also demonstrated that, as opposed to the recently reported results~\cite{Hashimoto17}, OTOC can be found to grow exponentially in chaotic systems when averaged appropriately.

Note that the level-spacing statistics is only one of the ways to study spectral correlations, and it only captures those at short ranges. Other statistical tools can uncover additional information hidden in the Lyapunovian. One interesting question is to study long-range correlations in the spectra of Lyapunov operators with such tools as spectral rigidity.

\begin{acknowledgments}
	This research was supported by NSF DMR-1613029, US-ARO (contract No. W911NF1310172), and Simons Foundation (E.B.R. and V.G.). The authors are grateful to Edward Ott, Leonid Bunimovich, and Shmuel Fishman for insightful discussions. The authors acknowledge the University of Maryland supercomputing resources (http://hpcc.umd.edu) made available for conducting the research reported in this paper. \vspace{-10pt}
\end{acknowledgments}


\begin{thebibliography}{67}%
	\makeatletter
	\providecommand \@ifxundefined [1]{%
		\@ifx{#1\undefined}
	}%
	\providecommand \@ifnum [1]{%
		\ifnum #1\expandafter \@firstoftwo
		\else \expandafter \@secondoftwo
		\fi
	}%
	\providecommand \@ifx [1]{%
		\ifx #1\expandafter \@firstoftwo
		\else \expandafter \@secondoftwo
		\fi
	}%
	\providecommand \natexlab [1]{#1}%
	\providecommand \enquote  [1]{``#1''}%
	\providecommand \bibnamefont  [1]{#1}%
	\providecommand \bibfnamefont [1]{#1}%
	\providecommand \citenamefont [1]{#1}%
	\providecommand \href@noop [0]{\@secondoftwo}%
	\providecommand \href [0]{\begingroup \@sanitize@url \@href}%
	\providecommand \@href[1]{\@@startlink{#1}\@@href}%
	\providecommand \@@href[1]{\endgroup#1\@@endlink}%
	\providecommand \@sanitize@url [0]{\catcode `\\12\catcode `\$12\catcode
		`\&12\catcode `\#12\catcode `\^12\catcode `\_12\catcode `\%12\relax}%
	\providecommand \@@startlink[1]{}%
	\providecommand \@@endlink[0]{}%
	\providecommand \url  [0]{\begingroup\@sanitize@url \@url }%
	\providecommand \@url [1]{\endgroup\@href {#1}{\urlprefix }}%
	\providecommand \urlprefix  [0]{URL }%
	\providecommand \Eprint [0]{\href }%
	\providecommand \doibase [0]{https://doi.org/}%
	\providecommand \selectlanguage [0]{\@gobble}%
	\providecommand \bibinfo  [0]{\@secondoftwo}%
	\providecommand \bibfield  [0]{\@secondoftwo}%
	\providecommand \translation [1]{[#1]}%
	\providecommand \BibitemOpen [0]{}%
	\providecommand \bibitemStop [0]{}%
	\providecommand \bibitemNoStop [0]{.\EOS\space}%
	\providecommand \EOS [0]{\spacefactor3000\relax}%
	\providecommand \BibitemShut  [1]{\csname bibitem#1\endcsname}%
	\let\auto@bib@innerbib\@empty
	\bibitem [{\citenamefont {Bohigas}\ \emph {et~al.}(1984)\citenamefont
		{Bohigas}, \citenamefont {Giannoni},\ and\ \citenamefont
		{Schmit}}]{Bohigas84}%
	\BibitemOpen
	\bibfield  {author} {\bibinfo {author} {\bibfnamefont {O.}~\bibnamefont
			{Bohigas}}, \bibinfo {author} {\bibfnamefont {M.~J.}\ \bibnamefont
			{Giannoni}},\ and\ \bibinfo {author} {\bibfnamefont {C.}~\bibnamefont
			{Schmit}},\ }\href {https://doi.org/10.1103/PhysRevLett.52.1} {\bibfield
		{journal} {\bibinfo  {journal} {Phys. Rev. Lett.}\ }\textbf {\bibinfo
			{volume} {52}},\ \bibinfo {pages} {1} (\bibinfo {year} {1984})}\BibitemShut
	{NoStop}%
	\bibitem [{\citenamefont {Casati}\ \emph {et~al.}(1980)\citenamefont {Casati},
		\citenamefont {Valz-Gris},\ and\ \citenamefont {Guarnieri}}]{Casati80}%
	\BibitemOpen
	\bibfield  {author} {\bibinfo {author} {\bibfnamefont {G.}~\bibnamefont
			{Casati}}, \bibinfo {author} {\bibfnamefont {F.}~\bibnamefont {Valz-Gris}},\
		and\ \bibinfo {author} {\bibfnamefont {I.}~\bibnamefont {Guarnieri}},\ }\href
	{https://doi.org/10.1007/BF02798790} {\bibfield  {journal} {\bibinfo
			{journal} {Lett. Nuovo Cimento}\ }\textbf {\bibinfo {volume} {28}},\ \bibinfo
		{pages} {279} (\bibinfo {year} {1980})}\BibitemShut {NoStop}%
	\bibitem [{\citenamefont {Gutzwiller}(1991)}]{gutzwiller1991chaos}%
	\BibitemOpen
	\bibfield  {author} {\bibinfo {author} {\bibfnamefont {M.~C.}\ \bibnamefont
			{Gutzwiller}},\ }\href {http://www.springer.com/us/book/9780387971735} {\emph
		{\bibinfo {title} {Chaos in Classical and Quantum Mechanics}}},\
	Interdisciplinary Applied Mathematics\ (\bibinfo  {publisher} {Springer New
		York},\ \bibinfo {year} {1991})\BibitemShut {NoStop}%
	\bibitem [{\citenamefont {Berry}(1985)}]{berry1985semiclassical}%
	\BibitemOpen
	\bibfield  {author} {\bibinfo {author} {\bibfnamefont {M.~V.}\ \bibnamefont
			{Berry}},\ }\href {https://doi.org/10.1098/rspa.1985.0078} {\bibfield
		{journal} {\bibinfo  {journal} {Proc. R. Soc. London, Ser. A}\ }\textbf {\bibinfo
			{volume} {400}},\ \bibinfo {pages} {229} (\bibinfo {year}
		{1985})}\BibitemShut {NoStop}%
	\bibitem [{\citenamefont {Andreev}\ \emph
		{et~al.}(1996{\natexlab{a}})\citenamefont {Andreev}, \citenamefont {Agam},
		\citenamefont {Simons},\ and\ \citenamefont
		{Altshuler}}]{andreev1996quantum}%
	\BibitemOpen
	\bibfield  {author} {\bibinfo {author} {\bibfnamefont {A.~V.}\ \bibnamefont
			{Andreev}}, \bibinfo {author} {\bibfnamefont {O.}~\bibnamefont {Agam}},
		\bibinfo {author} {\bibfnamefont {B.~D.}\ \bibnamefont {Simons}},\ and\
		\bibinfo {author} {\bibfnamefont {B.~L.}\ \bibnamefont {Altshuler}},\ }\href
	{https://doi.org/10.1103/PhysRevLett.76.3947} {\bibfield  {journal} {\bibinfo
			{journal} {Phys. Rev. Lett.}\ }\textbf {\bibinfo {volume} {76}},\ \bibinfo
		{pages} {3947} (\bibinfo {year} {1996}{\natexlab{a}})}\BibitemShut {NoStop}%
	\bibitem [{\citenamefont {Andreev}\ \emph
		{et~al.}(1996{\natexlab{b}})\citenamefont {Andreev}, \citenamefont {Simons},
		\citenamefont {Agam},\ and\ \citenamefont
		{Altshuler}}]{andreev1996semiclassical}%
	\BibitemOpen
	\bibfield  {author} {\bibinfo {author} {\bibfnamefont {A.~V.}\ \bibnamefont
			{Andreev}}, \bibinfo {author} {\bibfnamefont {B.~D.}\ \bibnamefont {Simons}},
		\bibinfo {author} {\bibfnamefont {O.}~\bibnamefont {Agam}},\ and\ \bibinfo
		{author} {\bibfnamefont {B.~L.}\ \bibnamefont {Altshuler}},\ }\href
	{https://doi.org/10.1016/S0550-3213(96)00473-7} {\bibfield  {journal}
		{\bibinfo  {journal} {Nucl. Phys. B}\ }\textbf {\bibinfo {volume} {482}},\
		\bibinfo {pages} {536 } (\bibinfo {year} {1996}{\natexlab{b}})}\BibitemShut
	{NoStop}%
	\bibitem [{\citenamefont {Altland}\ \emph {et~al.}(2015)\citenamefont
		{Altland}, \citenamefont {Gnutzmann}, \citenamefont {Haake},\ and\
		\citenamefont {Micklitz}}]{altland2015review}%
	\BibitemOpen
	\bibfield  {author} {\bibinfo {author} {\bibfnamefont {A.}~\bibnamefont
			{Altland}}, \bibinfo {author} {\bibfnamefont {S.}~\bibnamefont {Gnutzmann}},
		\bibinfo {author} {\bibfnamefont {F.}~\bibnamefont {Haake}},\ and\ \bibinfo
		{author} {\bibfnamefont {T.}~\bibnamefont {Micklitz}},\ }\href
	{http://stacks.iop.org/0034-4885/78/i=8/a=086001} {\bibfield  {journal}
		{\bibinfo  {journal} {Rep. Prog. Phys.}\ }\textbf {\bibinfo {volume} {78}},\
		\bibinfo {pages} {086001} (\bibinfo {year} {2015})}\BibitemShut {NoStop}%
	\bibitem [{\citenamefont {Stechel}\ and\ \citenamefont
		{Heller}(1984)}]{Stechel84}%
	\BibitemOpen
	\bibfield  {author} {\bibinfo {author} {\bibfnamefont {E.~B.}\ \bibnamefont
			{Stechel}}\ and\ \bibinfo {author} {\bibfnamefont {E.~J.}\ \bibnamefont
			{Heller}},\ }\href {https://doi.org/10.1146/annurev.pc.35.100184.003023}
	{\bibfield  {journal} {\bibinfo  {journal} {Annu. Rev. Phys. Chern.}\ }\textbf
		{\bibinfo {volume} {35}},\ \bibinfo {pages} {563} (\bibinfo {year}
		{1984})}\BibitemShut {NoStop}%
	\bibitem [{\citenamefont {Zelditch}(2005)}]{Zelditch05}%
	\BibitemOpen
	\bibfield  {author} {\bibinfo {author} {\bibfnamefont {S.}~\bibnamefont
			{Zelditch}},\ }\href {https://arxiv.org/abs/math-ph/0503026} {\bibfield
		{journal} {\bibinfo  {journal} {arXiv:math-ph/0503026}\ } (\bibinfo {year}
		{2005})}\BibitemShut {NoStop}%
	\bibitem [{\citenamefont {Zelditch}(2010)}]{Zelditch10}%
	\BibitemOpen
	\bibfield  {author} {\bibinfo {author} {\bibfnamefont {S.}~\bibnamefont
			{Zelditch}},\ }\href {https://projecteuclid.org:443/euclid.cdm/1335196624}
	{\bibfield  {journal} {\bibinfo  {journal} {Current Developments in
				Mathematics}\ }\textbf {\bibinfo {volume} {2009}},\ \bibinfo {pages} {115}
		(\bibinfo {year} {2010})}\BibitemShut {NoStop}%
	\bibitem [{\citenamefont {Berry}(1977)}]{Berry77}%
	\BibitemOpen
	\bibfield  {author} {\bibinfo {author} {\bibfnamefont {M.~V.}\ \bibnamefont
			{Berry}},\ }\href {http://stacks.iop.org/0305-4470/10/i=12/a=016} {\bibfield
		{journal} {\bibinfo  {journal} {J Phys A}\ }\textbf {\bibinfo {volume}
			{10}},\ \bibinfo {pages} {2083} (\bibinfo {year} {1977})}\BibitemShut
	{NoStop}%
	\bibitem [{\citenamefont {Berry}(2002)}]{Berry02}%
	\BibitemOpen
	\bibfield  {author} {\bibinfo {author} {\bibfnamefont {M.~V.}\ \bibnamefont
			{Berry}},\ }\href {http://stacks.iop.org/0305-4470/35/i=13/a=301} {\bibfield
		{journal} {\bibinfo  {journal} {J Phys A}\ }\textbf {\bibinfo {volume}
			{35}},\ \bibinfo {pages} {3025} (\bibinfo {year} {2002})}\BibitemShut
	{NoStop}%
	\bibitem [{\citenamefont {Bl\"umel}\ and\ \citenamefont
		{Smilansky}(1988)}]{Smilansky88}%
	\BibitemOpen
	\bibfield  {author} {\bibinfo {author} {\bibfnamefont {R.}~\bibnamefont
			{Bl\"umel}}\ and\ \bibinfo {author} {\bibfnamefont {U.}~\bibnamefont
			{Smilansky}},\ }\href {https://doi.org/10.1103/PhysRevLett.60.477} {\bibfield
		{journal} {\bibinfo  {journal} {Phys. Rev. Lett.}\ }\textbf {\bibinfo
			{volume} {60}},\ \bibinfo {pages} {477} (\bibinfo {year} {1988})}\BibitemShut
	{NoStop}%
	\bibitem [{\citenamefont {Casati}\ \emph {et~al.}(1990)\citenamefont {Casati},
		\citenamefont {Guarneri},\ and\ \citenamefont {Shepelyansky}}]{Casati90}%
	\BibitemOpen
	\bibfield  {author} {\bibinfo {author} {\bibfnamefont {G.}~\bibnamefont
			{Casati}}, \bibinfo {author} {\bibfnamefont {I.}~\bibnamefont {Guarneri}},\
		and\ \bibinfo {author} {\bibfnamefont {D.}~\bibnamefont {Shepelyansky}},\
	}\href {https://doi.org/10.1016/0378-4371(90)90330-U} {\bibfield  {journal}
		{\bibinfo  {journal} {Physica A}\ }\textbf {\bibinfo {volume} {163}},\
		\bibinfo {pages} {205 } (\bibinfo {year} {1990})}\BibitemShut {NoStop}%
	\bibitem [{\citenamefont {Peres}(1984)}]{Peres84}%
	\BibitemOpen
	\bibfield  {author} {\bibinfo {author} {\bibfnamefont {A.}~\bibnamefont
			{Peres}},\ }\href {https://doi.org/10.1103/PhysRevA.30.1610} {\bibfield
		{journal} {\bibinfo  {journal} {Phys. Rev. A}\ }\textbf {\bibinfo {volume}
			{30}},\ \bibinfo {pages} {1610} (\bibinfo {year} {1984})}\BibitemShut
	{NoStop}%
	\bibitem [{\citenamefont {Peres}(1995)}]{Peres95}%
	\BibitemOpen
	\bibfield  {author} {\bibinfo {author} {\bibfnamefont {A.}~\bibnamefont
			{Peres}},\ }\href@noop {} {\emph {\bibinfo {title} {Quantum Theory: Concepts
				and Methods}}}\ (\bibinfo  {publisher} {Kluwer, Dordrecht},\ \bibinfo {year}
	{1995})\BibitemShut {NoStop}%
	\bibitem [{\citenamefont {Wisniacki}\ \emph {et~al.}(2002)\citenamefont
		{Wisniacki}, \citenamefont {Vergini}, \citenamefont {Pastawski},\ and\
		\citenamefont {Cucchietti}}]{Wisniacki02}%
	\BibitemOpen
	\bibfield  {author} {\bibinfo {author} {\bibfnamefont {D.~A.}\ \bibnamefont
			{Wisniacki}}, \bibinfo {author} {\bibfnamefont {E.~G.}\ \bibnamefont
			{Vergini}}, \bibinfo {author} {\bibfnamefont {H.~M.}\ \bibnamefont
			{Pastawski}},\ and\ \bibinfo {author} {\bibfnamefont {F.~M.}\ \bibnamefont
			{Cucchietti}},\ }\href {https://doi.org/10.1103/PhysRevE.65.055206}
	{\bibfield  {journal} {\bibinfo  {journal} {Phys. Rev. E}\ }\textbf {\bibinfo
			{volume} {65}},\ \bibinfo {pages} {055206(R)} (\bibinfo {year}
		{2002})}\BibitemShut {NoStop}%
	\bibitem [{\citenamefont {Cucchietti}\ \emph {et~al.}(2002)\citenamefont
		{Cucchietti}, \citenamefont {Lewenkopf}, \citenamefont {Mucciolo},
		\citenamefont {Pastawski},\ and\ \citenamefont {Vallejos}}]{Cucchietti02}%
	\BibitemOpen
	\bibfield  {author} {\bibinfo {author} {\bibfnamefont {F.~M.}\ \bibnamefont
			{Cucchietti}}, \bibinfo {author} {\bibfnamefont {C.~H.}\ \bibnamefont
			{Lewenkopf}}, \bibinfo {author} {\bibfnamefont {E.~R.}\ \bibnamefont
			{Mucciolo}}, \bibinfo {author} {\bibfnamefont {H.~M.}\ \bibnamefont
			{Pastawski}},\ and\ \bibinfo {author} {\bibfnamefont {R.~O.}\ \bibnamefont
			{Vallejos}},\ }\href {https://doi.org/10.1103/PhysRevE.65.046209} {\bibfield
		{journal} {\bibinfo  {journal} {Phys. Rev. E}\ }\textbf {\bibinfo {volume}
			{65}},\ \bibinfo {pages} {046209} (\bibinfo {year} {2002})}\BibitemShut
	{NoStop}%
	\bibitem [{\citenamefont {Rozenbaum}\ \emph {et~al.}(2017)\citenamefont
		{Rozenbaum}, \citenamefont {Ganeshan},\ and\ \citenamefont
		{Galitski}}]{Rozenbaum17}%
	\BibitemOpen
	\bibfield  {author} {\bibinfo {author} {\bibfnamefont {E.~B.}\ \bibnamefont
			{Rozenbaum}}, \bibinfo {author} {\bibfnamefont {S.}~\bibnamefont
			{Ganeshan}},\ and\ \bibinfo {author} {\bibfnamefont {V.}~\bibnamefont
			{Galitski}},\ }\href {https://doi.org/10.1103/PhysRevLett.118.086801}
	{\bibfield  {journal} {\bibinfo  {journal} {Phys. Rev. Lett.}\ }\textbf
		{\bibinfo {volume} {118}},\ \bibinfo {pages} {086801} (\bibinfo {year}
		{2017})}\BibitemShut {NoStop}%
	\bibitem [{\citenamefont {Kitaev}(2014)}]{kitaev}%
	\BibitemOpen
	\bibfield  {author} {\bibinfo {author} {\bibfnamefont {A.}~\bibnamefont
			{Kitaev}},\ }\href {http://online.kitp.ucsb.edu/online/entangled15/kitaev/}
	{\bibfield  {journal} {\bibinfo  {journal} {KITP,
				http://online.kitp.ucsb.edu/online/ entangled15/kitaev/}\ } (\bibinfo {year}
		{2014})}\BibitemShut {NoStop}%
	\bibitem [{\citenamefont {Maldacena}\ \emph {et~al.}(2016)\citenamefont
		{Maldacena}, \citenamefont {Shenker},\ and\ \citenamefont
		{Stanford}}]{Maldacena16}%
	\BibitemOpen
	\bibfield  {author} {\bibinfo {author} {\bibfnamefont {J.}~\bibnamefont
			{Maldacena}}, \bibinfo {author} {\bibfnamefont {S.~H.}\ \bibnamefont
			{Shenker}},\ and\ \bibinfo {author} {\bibfnamefont {D.}~\bibnamefont
			{Stanford}},\ }\href {https://doi.org/10.1007/JHEP08(2016)106} {\bibfield
		{journal} {\bibinfo  {journal} {J. High Energy Phys.}\ }\textbf {\bibinfo
			{volume} {2016:106}}\bibinfo  {number} { (8)},\ \bibinfo {pages}
		{1}}\BibitemShut {NoStop}%
	\bibitem [{\citenamefont {Swingle}\ \emph {et~al.}(2016)\citenamefont
		{Swingle}, \citenamefont {Bentsen}, \citenamefont {Schleier-Smith},\ and\
		\citenamefont {Hayden}}]{swingle2016measuring}%
	\BibitemOpen
	\bibfield  {number} {  }\bibfield  {author} {\bibinfo {author} {\bibfnamefont
			{B.}~\bibnamefont {Swingle}}, \bibinfo {author} {\bibfnamefont
			{G.}~\bibnamefont {Bentsen}}, \bibinfo {author} {\bibfnamefont
			{M.}~\bibnamefont {Schleier-Smith}},\ and\ \bibinfo {author} {\bibfnamefont
			{P.}~\bibnamefont {Hayden}},\ }\href
	{https://doi.org/10.1103/PhysRevA.94.040302} {\bibfield  {journal} {\bibinfo
			{journal} {Phys. Rev. A}\ }\textbf {\bibinfo {volume} {94}},\ \bibinfo
		{pages} {040302(R)} (\bibinfo {year} {2016})}\BibitemShut {NoStop}%
	\bibitem [{\citenamefont {Yao}\ \emph {et~al.}(2016)\citenamefont {Yao},
		\citenamefont {Grusdt}, \citenamefont {Swingle}, \citenamefont {Lukin},
		\citenamefont {Stamper-Kurn}, \citenamefont {Moore},\ and\ \citenamefont
		{Demler}}]{yao2016interferometric}%
	\BibitemOpen
	\bibfield  {author} {\bibinfo {author} {\bibfnamefont {N.~Y.}\ \bibnamefont
			{Yao}}, \bibinfo {author} {\bibfnamefont {F.}~\bibnamefont {Grusdt}},
		\bibinfo {author} {\bibfnamefont {B.}~\bibnamefont {Swingle}}, \bibinfo
		{author} {\bibfnamefont {M.~D.}\ \bibnamefont {Lukin}}, \bibinfo {author}
		{\bibfnamefont {D.~M.}\ \bibnamefont {Stamper-Kurn}}, \bibinfo {author}
		{\bibfnamefont {J.~E.}\ \bibnamefont {Moore}},\ and\ \bibinfo {author}
		{\bibfnamefont {E.~A.}\ \bibnamefont {Demler}},\ }\href
	{https://arxiv.org/abs/1607.01801} {\bibfield  {journal} {\bibinfo  {journal}
			{{\tt arXiv:1607.01801}}\ } (\bibinfo {year} {2016})}\BibitemShut {NoStop}%
	\bibitem [{\citenamefont {Huang}\ \emph {et~al.}(2017)\citenamefont {Huang},
		\citenamefont {Zhang},\ and\ \citenamefont {Chen}}]{huang2016out}%
	\BibitemOpen
	\bibfield  {author} {\bibinfo {author} {\bibfnamefont {Y.}~\bibnamefont
			{Huang}}, \bibinfo {author} {\bibfnamefont {Y.-L.}\ \bibnamefont {Zhang}},\
		and\ \bibinfo {author} {\bibfnamefont {X.}~\bibnamefont {Chen}},\ }\href
	{https://doi.org/10.1002/andp.201600318} {\bibfield  {journal} {\bibinfo
			{journal} {Ann. Phys.}\ }\textbf {\bibinfo {volume} {529}},\ \bibinfo {pages}
		{1600318} (\bibinfo {year} {2017})}\BibitemShut {NoStop}%
	\bibitem [{\citenamefont {Fan}\ \emph {et~al.}(2017)\citenamefont {Fan},
		\citenamefont {Zhang}, \citenamefont {Shen},\ and\ \citenamefont
		{Zhai}}]{fan2016out}%
	\BibitemOpen
	\bibfield  {author} {\bibinfo {author} {\bibfnamefont {R.}~\bibnamefont
			{Fan}}, \bibinfo {author} {\bibfnamefont {P.}~\bibnamefont {Zhang}}, \bibinfo
		{author} {\bibfnamefont {H.}~\bibnamefont {Shen}},\ and\ \bibinfo {author}
		{\bibfnamefont {H.}~\bibnamefont {Zhai}},\ }\href
	{https://doi.org/10.1016/j.scib.2017.04.011} {\bibfield  {journal} {\bibinfo
			{journal} {Sci. Bull.}\ }\textbf {\bibinfo {volume} {62}},\ \bibinfo {pages}
		{707 } (\bibinfo {year} {2017})}\BibitemShut {NoStop}%
	\bibitem [{\citenamefont {Chen}(2016)}]{chen2016quantum}%
	\BibitemOpen
	\bibfield  {author} {\bibinfo {author} {\bibfnamefont {Y.}~\bibnamefont
			{Chen}},\ }\href {https://arxiv.org/abs/1608.02765} {\bibfield  {journal}
		{\bibinfo  {journal} {{\tt arXiv:1608.02765}}\ } (\bibinfo {year}
		{2016})}\BibitemShut {NoStop}%
	\bibitem [{\citenamefont {Swingle}\ and\ \citenamefont
		{Chowdhury}(2017)}]{swingle2016slow}%
	\BibitemOpen
	\bibfield  {author} {\bibinfo {author} {\bibfnamefont {B.}~\bibnamefont
			{Swingle}}\ and\ \bibinfo {author} {\bibfnamefont {D.}~\bibnamefont
			{Chowdhury}},\ }\href {https://doi.org/10.1103/PhysRevB.95.060201} {\bibfield
		{journal} {\bibinfo  {journal} {Phys. Rev. B}\ }\textbf {\bibinfo {volume}
			{95}},\ \bibinfo {pages} {060201(R)} (\bibinfo {year} {2017})}\BibitemShut
	{NoStop}%
	\bibitem [{\citenamefont {Syzranov}\ \emph {et~al.}(2019)\citenamefont
		{Syzranov}, \citenamefont {Gorshkov},\ and\ \citenamefont
		{Galitski}}]{Syzranov17}%
	\BibitemOpen
	\bibfield  {author} {\bibinfo {author} {\bibfnamefont {S.}~\bibnamefont
			{Syzranov}}, \bibinfo {author} {\bibfnamefont {A.}~\bibnamefont {Gorshkov}},\
		and\ \bibinfo {author} {\bibfnamefont {V.}~\bibnamefont {Galitski}},\ }\href
	{https://doi.org/10.1016/j.aop.2019.03.008} {\bibfield  {journal} {\bibinfo
			{journal} {Ann. Phys.}\ }\textbf {\bibinfo {volume} {405}},\ \bibinfo {pages}
		{1 } (\bibinfo {year} {2019})}\BibitemShut {NoStop}%
	\bibitem [{\citenamefont {Aleiner}\ and\ \citenamefont
		{Larkin}(1996)}]{aleiner1996divergence}%
	\BibitemOpen
	\bibfield  {author} {\bibinfo {author} {\bibfnamefont {I.~L.}\ \bibnamefont
			{Aleiner}}\ and\ \bibinfo {author} {\bibfnamefont {A.~I.}\ \bibnamefont
			{Larkin}},\ }\href {https://doi.org/10.1103/PhysRevB.54.14423} {\bibfield
		{journal} {\bibinfo  {journal} {Phys. Rev. B}\ }\textbf {\bibinfo {volume}
			{54}},\ \bibinfo {pages} {14423} (\bibinfo {year} {1996})}\BibitemShut
	{NoStop}%
	\bibitem [{\citenamefont {Aleiner}\ and\ \citenamefont
		{Larkin}(1997)}]{aleiner1997divergence_2}%
	\BibitemOpen
	\bibfield  {author} {\bibinfo {author} {\bibfnamefont {I.~L.}\ \bibnamefont
			{Aleiner}}\ and\ \bibinfo {author} {\bibfnamefont {A.~I.}\ \bibnamefont
			{Larkin}},\ }\href {https://doi.org/10.1103/PhysRevE.55.R1243} {\bibfield
		{journal} {\bibinfo  {journal} {Phys. Rev. E}\ }\textbf {\bibinfo {volume}
			{55}},\ \bibinfo {pages} {R1243} (\bibinfo {year} {1997})}\BibitemShut
	{NoStop}%
	\bibitem [{\citenamefont {Aleiner}\ \emph {et~al.}(2016)\citenamefont
		{Aleiner}, \citenamefont {Faoro},\ and\ \citenamefont
		{Ioffe}}]{aleiner2016microscopic}%
	\BibitemOpen
	\bibfield  {author} {\bibinfo {author} {\bibfnamefont {I.~L.}\ \bibnamefont
			{Aleiner}}, \bibinfo {author} {\bibfnamefont {L.}~\bibnamefont {Faoro}},\
		and\ \bibinfo {author} {\bibfnamefont {L.~B.}\ \bibnamefont {Ioffe}},\ }\href
	{https://doi.org/10.1016/j.aop.2016.09.006} {\bibfield  {journal} {\bibinfo
			{journal} {Ann. Phys.}\ }\textbf {\bibinfo {volume} {375}},\ \bibinfo {pages}
		{378 } (\bibinfo {year} {2016})}\BibitemShut {NoStop}%
	\bibitem [{\citenamefont {Sachdev}(2015)}]{Sachdev15}%
	\BibitemOpen
	\bibfield  {author} {\bibinfo {author} {\bibfnamefont {S.}~\bibnamefont
			{Sachdev}},\ }\href {https://doi.org/10.1103/PhysRevX.5.041025} {\bibfield
		{journal} {\bibinfo  {journal} {Phys. Rev. X}\ }\textbf {\bibinfo {volume}
			{5}},\ \bibinfo {pages} {041025} (\bibinfo {year} {2015})}\BibitemShut
	{NoStop}%
	\bibitem [{\citenamefont {Ara\'ujo~Lima}\ \emph {et~al.}(2013)\citenamefont
		{Ara\'ujo~Lima}, \citenamefont {Rodr\'{\i}guez-P\'erez},\ and\ \citenamefont
		{de~Aguiar}}]{Lima13}%
	\BibitemOpen
	\bibfield  {author} {\bibinfo {author} {\bibfnamefont {T.}~\bibnamefont
			{Ara\'ujo~Lima}}, \bibinfo {author} {\bibfnamefont {S.}~\bibnamefont
			{Rodr\'{\i}guez-P\'erez}},\ and\ \bibinfo {author} {\bibfnamefont {F.~M.}\
			\bibnamefont {de~Aguiar}},\ }\href
	{https://doi.org/10.1103/PhysRevE.87.062902} {\bibfield  {journal} {\bibinfo
			{journal} {Phys. Rev. E}\ }\textbf {\bibinfo {volume} {87}},\ \bibinfo
		{pages} {062902} (\bibinfo {year} {2013})}\BibitemShut {NoStop}%
	\bibitem [{\citenamefont {Larkin}\ and\ \citenamefont
		{Ovchinnikov}(1969)}]{Larkin69}%
	\BibitemOpen
	\bibfield  {author} {\bibinfo {author} {\bibfnamefont {A.}~\bibnamefont
			{Larkin}}\ and\ \bibinfo {author} {\bibfnamefont {{\relax Yu}.~N.}\
			\bibnamefont {Ovchinnikov}},\ } {\bibfield
		{journal} {\bibinfo  {journal} {Zh. Eksp. Teor. Fiz.}\ }\textbf {\bibinfo
			{volume} {55}},\ \bibinfo {pages} {2262} (\bibinfo {year} {1969})},\ \bibinfo
	{note} {[\href
		{http://www.jetp.ac.ru/cgi-bin/e/index/e/28/6/p1200?a=list}{Sov. Phys. -- JETP {\bf 28}, 1200 (1969)}]}\BibitemShut {NoStop}%
	\bibitem [{\citenamefont {Maldacena}\ and\ \citenamefont
		{Stanford}(2016)}]{maldacena2016remarks}%
	\BibitemOpen
	\bibfield  {author} {\bibinfo {author} {\bibfnamefont {J.}~\bibnamefont
			{Maldacena}}\ and\ \bibinfo {author} {\bibfnamefont {D.}~\bibnamefont
			{Stanford}},\ }\href {https://doi.org/10.1103/PhysRevD.94.106002} {\bibfield
		{journal} {\bibinfo  {journal} {Phys. Rev. D}\ }\textbf {\bibinfo {volume}
			{94}},\ \bibinfo {pages} {106002} (\bibinfo {year} {2016})}\BibitemShut
	{NoStop}%
	\bibitem [{\citenamefont {Bagrets}\ \emph {et~al.}(2016)\citenamefont
		{Bagrets}, \citenamefont {Altland},\ and\ \citenamefont
		{Kamenev}}]{bagrets2016sachdev}%
	\BibitemOpen
	\bibfield  {author} {\bibinfo {author} {\bibfnamefont {D.}~\bibnamefont
			{Bagrets}}, \bibinfo {author} {\bibfnamefont {A.}~\bibnamefont {Altland}},\
		and\ \bibinfo {author} {\bibfnamefont {A.}~\bibnamefont {Kamenev}},\ }\href
	{https://doi.org/10.1016/j.nuclphysb.2016.08.002} {\bibfield  {journal}
		{\bibinfo  {journal} {Nucl. Phys. B}\ }\textbf {\bibinfo {volume} {911}},\
		\bibinfo {pages} {191 } (\bibinfo {year} {2016})}\BibitemShut {NoStop}%
	\bibitem [{\citenamefont {Cotler}\ \emph {et~al.}(2017)\citenamefont {Cotler},
		\citenamefont {Hunter-Jones}, \citenamefont {Liu},\ and\ \citenamefont
		{Yoshida}}]{Cotler2017chaos}%
	\BibitemOpen
	\bibfield  {author} {\bibinfo {author} {\bibfnamefont {J.}~\bibnamefont
			{Cotler}}, \bibinfo {author} {\bibfnamefont {N.}~\bibnamefont
			{Hunter-Jones}}, \bibinfo {author} {\bibfnamefont {J.}~\bibnamefont {Liu}},\
		and\ \bibinfo {author} {\bibfnamefont {B.}~\bibnamefont {Yoshida}},\ }\href
	{https://doi.org/10.1007/JHEP11(2017)048} {\bibfield  {journal} {\bibinfo
			{journal} {J. High Energy Phys.}\ }\textbf {\bibinfo {volume} {2017}}\bibinfo
		{number} { (11)},\ \bibinfo {pages} {48}}\BibitemShut {NoStop}%
	\bibitem [{Note1()}]{Note1}%
	\BibitemOpen
	\bibfield  {number} {  }\bibinfo {note} {Note the qualitative difference
		between the finite-time spectrum of the single-particle Lyapunovian and the
		spectrum of infinite-time Lyapunov exponents in multidimensional classical
		models~\cite {GurAri16,*Hanada18}.}\BibitemShut {Stop}%
	\bibitem [{Par()}]{ParityComment}%
	\BibitemOpen
	\bibinfo {note} {Due to the reflection symmetries, the Hamiltonian can be
		written as a $4\times4$ block-diagonal matrix in a basis of functions with
		definite parities. As a result, there is no correlation (thus no repulsion)
		between the eigenvalues of different blocks. More details are given after
		Eq.~(\ref{SE}).}\BibitemShut {Stop}%
	\bibitem [{\citenamefont {Gr\"af}\ \emph {et~al.}(1992)\citenamefont {Gr\"af},
		\citenamefont {Harney}, \citenamefont {Lengeler}, \citenamefont {Lewenkopf},
		\citenamefont {Rangacharyulu}, \citenamefont {Richter}, \citenamefont
		{Schardt},\ and\ \citenamefont {Weidenm\"uller}}]{Graf92}%
	\BibitemOpen
	\bibfield  {author} {\bibinfo {author} {\bibfnamefont {{\relax
					H.-D}.}~\bibnamefont {Gr\"af}}, \bibinfo {author} {\bibfnamefont {H.~L.}\
			\bibnamefont {Harney}}, \bibinfo {author} {\bibfnamefont {H.}~\bibnamefont
			{Lengeler}}, \bibinfo {author} {\bibfnamefont {C.~H.}\ \bibnamefont
			{Lewenkopf}}, \bibinfo {author} {\bibfnamefont {C.}~\bibnamefont
			{Rangacharyulu}}, \bibinfo {author} {\bibfnamefont {A.}~\bibnamefont
			{Richter}}, \bibinfo {author} {\bibfnamefont {P.}~\bibnamefont {Schardt}},\
		and\ \bibinfo {author} {\bibfnamefont {H.~A.}\ \bibnamefont
			{Weidenm\"uller}},\ }\href {https://doi.org/10.1103/PhysRevLett.69.1296}
	{\bibfield  {journal} {\bibinfo  {journal} {Phys. Rev. Lett.}\ }\textbf
		{\bibinfo {volume} {69}},\ \bibinfo {pages} {1296} (\bibinfo {year}
		{1992})}\BibitemShut {NoStop}%
	\bibitem [{\citenamefont {Sieber}\ \emph {et~al.}(1993)\citenamefont {Sieber},
		\citenamefont {Smilansky}, \citenamefont {Creagh},\ and\ \citenamefont
		{Littlejohn}}]{Sieber93}%
	\BibitemOpen
	\bibfield  {author} {\bibinfo {author} {\bibfnamefont {M.}~\bibnamefont
			{Sieber}}, \bibinfo {author} {\bibfnamefont {U.}~\bibnamefont {Smilansky}},
		\bibinfo {author} {\bibfnamefont {S.~C.}\ \bibnamefont {Creagh}},\ and\
		\bibinfo {author} {\bibfnamefont {R.~G.}\ \bibnamefont {Littlejohn}},\ }\href
	{http://stacks.iop.org/0305-4470/26/i=22/a=022} {\bibfield  {journal}
		{\bibinfo  {journal} {J Phys A}\ }\textbf {\bibinfo {volume} {26}},\ \bibinfo
		{pages} {6217} (\bibinfo {year} {1993})}\BibitemShut {NoStop}%
	\bibitem [{\citenamefont {Alt}\ \emph {et~al.}(1999)\citenamefont {Alt},
		\citenamefont {Dembowski}, \citenamefont {Gr\"af}, \citenamefont
		{Hofferbert}, \citenamefont {Rehfeld}, \citenamefont {Richter},\ and\
		\citenamefont {Schmit}}]{Alt99}%
	\BibitemOpen
	\bibfield  {author} {\bibinfo {author} {\bibfnamefont {H.}~\bibnamefont
			{Alt}}, \bibinfo {author} {\bibfnamefont {C.}~\bibnamefont {Dembowski}},
		\bibinfo {author} {\bibfnamefont {{\relax H.-D}.}~\bibnamefont {Gr\"af}},
		\bibinfo {author} {\bibfnamefont {R.}~\bibnamefont {Hofferbert}}, \bibinfo
		{author} {\bibfnamefont {H.}~\bibnamefont {Rehfeld}}, \bibinfo {author}
		{\bibfnamefont {A.}~\bibnamefont {Richter}},\ and\ \bibinfo {author}
		{\bibfnamefont {C.}~\bibnamefont {Schmit}},\ }\href
	{https://doi.org/10.1103/PhysRevE.60.2851} {\bibfield  {journal} {\bibinfo
			{journal} {Phys. Rev. E}\ }\textbf {\bibinfo {volume} {60}},\ \bibinfo
		{pages} {2851} (\bibinfo {year} {1999})}\BibitemShut {NoStop}%
	\bibitem [{\citenamefont {Sinai}(1970)}]{Sinai70}%
	\BibitemOpen
	\bibfield  {author} {\bibinfo {author} {\bibfnamefont {Y.~G.}\ \bibnamefont
			{Sinai}},\ }\href {https://doi.org/10.1070/RM1970v025n02ABEH003794}
	{\bibfield  {journal} {\bibinfo  {journal} {Russian Math. Surv.}\ }\textbf
		{\bibinfo {volume} {25}},\ \bibinfo {pages} {137} (\bibinfo {year}
		{1970})}\BibitemShut {NoStop}%
	\bibitem [{\citenamefont {Bunimovich}(1974)}]{Bunimovich74}%
	\BibitemOpen
	\bibfield  {author} {\bibinfo {author} {\bibfnamefont {L.~A.}\ \bibnamefont
			{Bunimovich}},\ }\href {https://doi.org/10.1007/BF01075700} {\bibfield
		{journal} {\bibinfo  {journal} {Functional Analysis and Its Applications}\
		}\textbf {\bibinfo {volume} {8}},\ \bibinfo {pages} {254} (\bibinfo {year}
		{1974})}\BibitemShut {NoStop}%
	\bibitem [{\citenamefont {Bunimovich}(1979)}]{Bunimovich79}%
	\BibitemOpen
	\bibfield  {author} {\bibinfo {author} {\bibfnamefont {L.~A.}\ \bibnamefont
			{Bunimovich}},\ }\href {https://doi.org/10.1007/BF01197884} {\bibfield
		{journal} {\bibinfo  {journal} {Commun. Math. Phys.}\ }\textbf {\bibinfo
			{volume} {65}},\ \bibinfo {pages} {295} (\bibinfo {year} {1979})}\BibitemShut
	{NoStop}%
	\bibitem [{\citenamefont {Bunimovich}\ \emph {et~al.}(1991)\citenamefont
		{Bunimovich}, \citenamefont {Sinai},\ and\ \citenamefont
		{Chernov}}]{Bunimovich91}%
	\BibitemOpen
	\bibfield  {author} {\bibinfo {author} {\bibfnamefont {L.~A.}\ \bibnamefont
			{Bunimovich}}, \bibinfo {author} {\bibfnamefont {Y.~G.}\ \bibnamefont
			{Sinai}},\ and\ \bibinfo {author} {\bibfnamefont {N.~I.}\ \bibnamefont
			{Chernov}},\ }\href {https://doi.org/10.1070/RM1991v046n04ABEH002827}
	{\bibfield  {journal} {\bibinfo  {journal} {Russian Math. Surv.}\ }\textbf
		{\bibinfo {volume} {46}},\ \bibinfo {pages} {47} (\bibinfo {year}
		{1991})}\BibitemShut {NoStop}%
	\bibitem [{\citenamefont {Benettin}\ and\ \citenamefont
		{Strelcyn}(1978)}]{Benettin78}%
	\BibitemOpen
	\bibfield  {author} {\bibinfo {author} {\bibfnamefont {G.}~\bibnamefont
			{Benettin}}\ and\ \bibinfo {author} {\bibfnamefont {{\relax
					J.-M}.}~\bibnamefont {Strelcyn}},\ }\href
	{https://doi.org/10.1103/PhysRevA.17.773} {\bibfield  {journal} {\bibinfo
			{journal} {Phys. Rev. A}\ }\textbf {\bibinfo {volume} {17}},\ \bibinfo
		{pages} {773} (\bibinfo {year} {1978})}\BibitemShut {NoStop}%
	\bibitem [{\citenamefont {Dellago}\ and\ \citenamefont
		{Posch}(1995)}]{Dellago95}%
	\BibitemOpen
	\bibfield  {author} {\bibinfo {author} {\bibfnamefont {C.}~\bibnamefont
			{Dellago}}\ and\ \bibinfo {author} {\bibfnamefont {H.~A.}\ \bibnamefont
			{Posch}},\ }\href {https://doi.org/10.1103/PhysRevE.52.2401} {\bibfield
		{journal} {\bibinfo  {journal} {Phys. Rev. E}\ }\textbf {\bibinfo {volume}
			{52}},\ \bibinfo {pages} {2401} (\bibinfo {year} {1995})}\BibitemShut
	{NoStop}%
	\bibitem [{\citenamefont {Biham}\ and\ \citenamefont {Kvale}(1992)}]{Biham92}%
	\BibitemOpen
	\bibfield  {author} {\bibinfo {author} {\bibfnamefont {O.}~\bibnamefont
			{Biham}}\ and\ \bibinfo {author} {\bibfnamefont {M.}~\bibnamefont {Kvale}},\
	}\href {https://doi.org/10.1103/PhysRevA.46.6334} {\bibfield  {journal}
		{\bibinfo  {journal} {Phys. Rev. A}\ }\textbf {\bibinfo {volume} {46}},\
		\bibinfo {pages} {6334} (\bibinfo {year} {1992})}\BibitemShut {NoStop}%
	\bibitem [{\citenamefont {McDonald}\ and\ \citenamefont
		{Kaufman}(1979)}]{McDonald79}%
	\BibitemOpen
	\bibfield  {author} {\bibinfo {author} {\bibfnamefont {S.~W.}\ \bibnamefont
			{McDonald}}\ and\ \bibinfo {author} {\bibfnamefont {A.~N.}\ \bibnamefont
			{Kaufman}},\ }\href {https://doi.org/10.1103/PhysRevLett.42.1189} {\bibfield
		{journal} {\bibinfo  {journal} {Phys. Rev. Lett.}\ }\textbf {\bibinfo
			{volume} {42}},\ \bibinfo {pages} {1189} (\bibinfo {year}
		{1979})}\BibitemShut {NoStop}%
	\bibitem [{\citenamefont {Shudo}\ and\ \citenamefont
		{Shimizu}(1990)}]{Shudo90}%
	\BibitemOpen
	\bibfield  {author} {\bibinfo {author} {\bibfnamefont {A.}~\bibnamefont
			{Shudo}}\ and\ \bibinfo {author} {\bibfnamefont {Y.}~\bibnamefont
			{Shimizu}},\ }\href {https://doi.org/10.1103/PhysRevA.42.6264} {\bibfield
		{journal} {\bibinfo  {journal} {Phys. Rev. A}\ }\textbf {\bibinfo {volume}
			{42}},\ \bibinfo {pages} {6264} (\bibinfo {year} {1990})}\BibitemShut
	{NoStop}%
	\bibitem [{\citenamefont {Borgonovi}\ \emph {et~al.}(1996)\citenamefont
		{Borgonovi}, \citenamefont {Casati},\ and\ \citenamefont
		{Li}}]{Borgonovi96_1}%
	\BibitemOpen
	\bibfield  {author} {\bibinfo {author} {\bibfnamefont {F.}~\bibnamefont
			{Borgonovi}}, \bibinfo {author} {\bibfnamefont {G.}~\bibnamefont {Casati}},\
		and\ \bibinfo {author} {\bibfnamefont {B.}~\bibnamefont {Li}},\ }\href
	{https://doi.org/10.1103/PhysRevLett.77.4744} {\bibfield  {journal} {\bibinfo
			{journal} {Phys. Rev. Lett.}\ }\textbf {\bibinfo {volume} {77}},\ \bibinfo
		{pages} {4744} (\bibinfo {year} {1996})}\BibitemShut {NoStop}%
	\bibitem [{\citenamefont {Heuveline}(2003)}]{Heuveline03}%
	\BibitemOpen
	\bibfield  {author} {\bibinfo {author} {\bibfnamefont {V.}~\bibnamefont
			{Heuveline}},\ }\href {https://doi.org/10.1016/S0021-9991(02)00043-8}
	{\bibfield  {journal} {\bibinfo  {journal} {J. Comput. Phys.}\ }\textbf
		{\bibinfo {volume} {184}},\ \bibinfo {pages} {321 } (\bibinfo {year}
		{2003})}\BibitemShut {NoStop}%
	\bibitem [{\citenamefont {Baltes}\ and\ \citenamefont {Hilf}(1976)}]{Baltes78}%
	\BibitemOpen
	\bibfield  {author} {\bibinfo {author} {\bibfnamefont {H.~P.}\ \bibnamefont
			{Baltes}}\ and\ \bibinfo {author} {\bibfnamefont {E.~R.}\ \bibnamefont
			{Hilf}},\ }\href@noop {} {\emph {\bibinfo {title} {Spectra of finite
				systems}}}\ (\bibinfo  {publisher} {Mannheim: Bibliographisches Institut},\
	\bibinfo {year} {1976})\BibitemShut {NoStop}%
	\bibitem [{Wis()}]{WishartNote}%
	\BibitemOpen
	\bibinfo {note} {These matrices themselves do not belong to the Gaussian
		ensembles. In Ref.~\cite{AbulMagd09}, it is shown that for Wishart-ensemble
		matrices, the eigenvalue statistics of the bulk of the spectrum is still very
		well described by the universal Wigner-Dyson distributions. For operators
		that we consider, this appears to hold as well.}\BibitemShut {Stop}%
	\bibitem [{\citenamefont {Cotler}\ \emph {et~al.}(2018)\citenamefont {Cotler},
		\citenamefont {Ding},\ and\ \citenamefont {Penington}}]{cotler2017out}%
	\BibitemOpen
	\bibfield  {author} {\bibinfo {author} {\bibfnamefont {J.~S.}\ \bibnamefont
			{Cotler}}, \bibinfo {author} {\bibfnamefont {D.}~\bibnamefont {Ding}},\ and\
		\bibinfo {author} {\bibfnamefont {G.~R.}\ \bibnamefont {Penington}},\ }\href
	{https://doi.org/10.1016/j.aop.2018.07.020} {\bibfield  {journal} {\bibinfo
			{journal} {Ann. Phys.}\ }\textbf {\bibinfo {volume} {396}},\ \bibinfo {pages}
		{318 } (\bibinfo {year} {2018})}\BibitemShut {NoStop}%
	\bibitem [{\citenamefont {Moyal}(1949)}]{Moyal49}%
	\BibitemOpen
	\bibfield  {author} {\bibinfo {author} {\bibfnamefont {J.~E.}\ \bibnamefont
			{Moyal}},\ }\href {https://doi.org/10.1017/S0305004100000487} {\bibfield
		{journal} {\bibinfo  {journal} {Mathematical Proceedings of the Cambridge
				Philosophical Society}\ }\textbf {\bibinfo {volume} {45}},\ \bibinfo {pages}
		{99–124} (\bibinfo {year} {1949})}\BibitemShut {NoStop}%
	\bibitem [{\citenamefont {Zurek}\ and\ \citenamefont
		{Paz}(1994)}]{zurek1994decoherence}%
	\BibitemOpen
	\bibfield  {author} {\bibinfo {author} {\bibfnamefont {W.~H.}\ \bibnamefont
			{Zurek}}\ and\ \bibinfo {author} {\bibfnamefont {J.~P.}\ \bibnamefont
			{Paz}},\ }\href {https://doi.org/10.1103/PhysRevLett.72.2508} {\bibfield
		{journal} {\bibinfo  {journal} {Phys. Rev. Lett.}\ }\textbf {\bibinfo
			{volume} {72}},\ \bibinfo {pages} {2508} (\bibinfo {year}
		{1994})}\BibitemShut {NoStop}%
	\bibitem [{\citenamefont {Hashimoto}\ \emph {et~al.}(2017)\citenamefont
		{Hashimoto}, \citenamefont {Murata},\ and\ \citenamefont
		{Yoshii}}]{Hashimoto17}%
	\BibitemOpen
	\bibfield  {author} {\bibinfo {author} {\bibfnamefont {K.}~\bibnamefont
			{Hashimoto}}, \bibinfo {author} {\bibfnamefont {K.}~\bibnamefont {Murata}},\
		and\ \bibinfo {author} {\bibfnamefont {R.}~\bibnamefont {Yoshii}},\ }\href
	{https://doi.org/10.1007/JHEP10(2017)138} {\bibfield  {journal} {\bibinfo
			{journal} {J. High Energy Phys.}\ }\textbf {\bibinfo {volume} {2017}}\bibinfo
		{number} { (10)},\ \bibinfo {pages} {138}}\BibitemShut {NoStop}%
	\bibitem [{\citenamefont {Lozej}(2018)}]{Crt}%
	\BibitemOpen
	\bibfield  {number} {  }\bibfield  {author} {\bibinfo {author} {\bibfnamefont
			{{\relax {${\rm \check{C}}$}}.}~\bibnamefont {Lozej}},\ }\href@noop {}
	{}\bibinfo {howpublished} {Private Communication} (\bibinfo {year}
	{2018})\BibitemShut {NoStop}%
	\bibitem [{Pro()}]{ProbDist}%
	\BibitemOpen
	\bibinfo {note} {Note that the Wigner distribution $W_{\rm cl}(z)$ is
		everywhere positive, so that it is an actual probability distribution,
		reflecting the fact that the initial state does not have quantum
		interferences built in.}\BibitemShut {Stop}%
	\bibitem [{\citenamefont {Hunt}\ and\ \citenamefont {Ott}(2015)}]{Ott15}%
	\BibitemOpen
	\bibfield  {author} {\bibinfo {author} {\bibfnamefont {B.~R.}\ \bibnamefont
			{Hunt}}\ and\ \bibinfo {author} {\bibfnamefont {E.}~\bibnamefont {Ott}},\
	}\href {https://doi.org/10.1063/1.4922973} {\bibfield  {journal} {\bibinfo
			{journal} {Chaos: An Interdisciplinary Journal of Nonlinear Science}\
		}\textbf {\bibinfo {volume} {25}},\ \bibinfo {pages} {097618} (\bibinfo
		{year} {2015})}\BibitemShut {NoStop}%
	\bibitem [{LE_()}]{LE_Comment}%
	\BibitemOpen
	\bibinfo {note} {In Refs.~\cite{Benettin78,Dellago95,Biham92}, the billiard
		area $A=\pi+4$, hence they provide a different numerical value for the
		exponent ($0.43$) that scales as $A^{-1/2}$ and gives $1.15$ for
		$A=1$.}\BibitemShut {Stop}%
	\bibitem [{Note2()}]{Note2}%
	\BibitemOpen
	\bibinfo {note} {Note that classical Lyapunov exponent was first extracted in
		a related way in Refs.~\cite {Wisniacki02,*Cucchietti02} from Loschmidt echo,
		which is closely related to OTOC.}\BibitemShut {Stop}%
	\bibitem [{\citenamefont {Gur-Ari}\ \emph {et~al.}(2016)\citenamefont
		{Gur-Ari}, \citenamefont {Hanada},\ and\ \citenamefont {Shenker}}]{GurAri16}%
	\BibitemOpen
	\bibfield  {author} {\bibinfo {author} {\bibfnamefont {G.}~\bibnamefont
			{Gur-Ari}}, \bibinfo {author} {\bibfnamefont {M.}~\bibnamefont {Hanada}},\
		and\ \bibinfo {author} {\bibfnamefont {S.~H.}\ \bibnamefont {Shenker}},\
	}\href {https://doi.org/10.1007/JHEP02(2016)091} {\bibfield  {journal}
		{\bibinfo  {journal} {Journal of High Energy Physics}\ }\textbf {\bibinfo
			{volume} {2016}},\ \bibinfo {pages} {91} (\bibinfo {year}
		{2016})}\BibitemShut {NoStop}%
	\bibitem [{\citenamefont {Hanada}\ \emph {et~al.}(2018)\citenamefont {Hanada},
		\citenamefont {Shimada},\ and\ \citenamefont {Tezuka}}]{Hanada18}%
	\BibitemOpen
	\bibfield  {author} {\bibinfo {author} {\bibfnamefont {M.}~\bibnamefont
			{Hanada}}, \bibinfo {author} {\bibfnamefont {H.}~\bibnamefont {Shimada}},\
		and\ \bibinfo {author} {\bibfnamefont {M.}~\bibnamefont {Tezuka}},\ }\href
	{https://doi.org/10.1103/PhysRevE.97.022224} {\bibfield  {journal} {\bibinfo
			{journal} {Phys. Rev. E}\ }\textbf {\bibinfo {volume} {97}},\ \bibinfo
		{pages} {022224} (\bibinfo {year} {2018})}\BibitemShut {NoStop}%
	\bibitem [{\citenamefont {Abul-Magd}\ \emph {et~al.}(2009)\citenamefont
		{Abul-Magd}, \citenamefont {Akemann},\ and\ \citenamefont
		{Vivo}}]{AbulMagd09}%
	\BibitemOpen
	\bibfield  {author} {\bibinfo {author} {\bibfnamefont {A.~Y.}\ \bibnamefont
			{Abul-Magd}}, \bibinfo {author} {\bibfnamefont {G.}~\bibnamefont {Akemann}},\
		and\ \bibinfo {author} {\bibfnamefont {P.}~\bibnamefont {Vivo}},\ }\href
	{http://stacks.iop.org/1751-8121/42/i=17/a=175207} {\bibfield  {journal}
		{\bibinfo  {journal} {J. Phys. A: Math. Theor.}\ }\textbf {\bibinfo {volume}
			{42}},\ \bibinfo {pages} {175207} (\bibinfo {year} {2009})}\BibitemShut
	{NoStop}%
\end{thebibliography}

%

\end{document}